\documentclass[reprint,aps,prx,twocolumn,english,showpacs,superscriptaddress,amssymb,amsfonts]{revtex4-2}

\usepackage{graphicx}
\usepackage{epsfig}
\usepackage{color}
\usepackage{amsmath}
\usepackage{amsfonts}
\usepackage{mathrsfs}
\usepackage{txfonts}
\usepackage{color}
\usepackage{mathtools}

\usepackage[dvipsnames,svgnames,x11names,hyperref]{xcolor}

\usepackage[breaklinks]{hyperref}
\hypersetup{colorlinks=true, linkcolor=blue, citecolor=blue, filecolor=blue, urlcolor=blue}

\usepackage[normalem]{ulem}

\begin{document}
	\title{Discrete time crystals enforced by Floquet-Bloch scars}
	\author{Biao Huang}
	\email{phys.huang.biao@gmail.com}
	\affiliation{Kavli Institute for Theoretical Sciences, University of Chinese Academy of Sciences, Beijing 100190, China}
	\author{Tsz-Him Leung}
	\affiliation{Department of Physics, University of California, Berkeley, USA}
	\author{Dan Stamper-Kurn}
	\affiliation{Department of Physics, University of California, Berkeley, USA}
	\affiliation{Materials Sciences Division, Lawrence Berkeley National Laboratory, USA}
	\author{W. Vincent Liu}
	\email{wvliu@pitt.edu}
	\affiliation{Department of Physics and Astronomy, University of Pittsburgh, Pittsburgh, PA 15260, USA}
	\affiliation{Shenzhen Institute for Quantum Science and Engineering, Southern University of Science and Technology, Shenzhen 518055, China}
	\date{\today}
	\begin{abstract}
		We analytically identify a new class of quantum scars protected by spatiotemporal translation symmetries, dubbed Floquet-Bloch scars. They distinguish from previous (quasi-)static scars by a rigid spectral pairing only possible in Floquet systems, where strong interaction and  drivings equalize the quasienergy corrections to all scars and maintain their spectral spacings against generic bilinear perturbations. Scars then enforce the spatial localization and rigid discrete time crystal (DTC) oscillations as verified numerically in a trimerized kagome lattice model relevant to recent cold atom experiments. 
		Our analytical solutions offer a potential scheme to understand the mechanisms for more generic translation-invariant DTCs.
	\end{abstract}
	\maketitle

	{\noindent\em\color{blue}Introduction} --- Systems far from equilibrium have become a fertile ground cultivating unexpected phenomena recently. Among them, discrete time crystals (DTC)~\cite{Khemani2016,Else2016,Yao2017,Ho2017,Sacha2015,Zhang2017,Choi2017} constitute an intriguing example. As foundational concepts of ground state and temperature fall apart in the absence of thermal equilibrium, Landau's theory of symmetry breaking~\cite{LandauStat} is replaced by new principles like spectral pairing and eigenstate orders~\cite{Khemani2016,Else2016} in handling time translation symmetries. That results in the DTC phenomena where Hamiltonians $ \hat{H}(t+T) = \hat{H}(t) $ give rise to observables $ O(t+NT)=O(t) $ ($ 1<N\in\mathbb{Z} $) oscillating like a temporal charge/spin density wave. Crucially, the periodicity $ NT $ demands no fine-tuning and withstands generic perturbations.
	
	The concept of DTCs has been considered in several physical realizations~\cite{Rovny2018,Pal2018,Mi2022,Randall2021,Kyprianidis2021,Estarellas2020,Frey2022}.
	While the strongly disordered cases are relatively well understood~\cite{Khemani2016,Else2016,Yao2017,Keyserlingk2016}, the possibility of DTCs in translation-invariant ordered systems is less clear.  Empirical evidence for DTCs is accumulating in both physical and numerical experiments~\cite{Rovny2018,Pal2018,Huang2018,Russomanno2017,Zeng2017,Lyu2020,Yu2019,Mizuta2018,Barfknecht2019}. However, analytical explanations based on many-body localization (MBL)~\cite{Abanin2019} or prethermalization~\cite{Else2017} do not seem to apply to these cases. Recently, it was indicated that quasi-conservation laws~\cite{Luitz2020,Ho2020}, which can be enhanced by single-particle terms, may help protect phenomena pertinent to DTCs. Meanwhile, the initial state dependence of clean DTCs~\cite{Khemani2019b,Luitz2020} has been reexamined in terms of scar physics~\cite{Turner2018} in recent numerics~\cite{Pizzi2020,Yarloo2020}. Altogether, continued investigation on DTCs in non-disordered systems, with the objectives of uncovering the underlying mechanism that supports the DTC and the specific role of many-body (vs.\ single-particle) effects, is warranted.


	
	In this Letter, we gain insights on these two research objectives by studying a small cluster of soft-core bosons on driven trimerized kagome lattices, relevant to recent experiments~\cite{Barter2020} and feasible for numerical verifications. We find analytically that it is a class of quantum scars {\em protected by} spatiotemporal translation invariance, dubbed Floquet-Bloch scars (FBS), that gives rise to DTC behaviors for sublattice density oscillations. 
	FBS's identified here {\em neither} exploit a static scar (i.e. ``PXP" model~\cite{Turner2018,Khemani2019a,Choi2019}) {\em nor} end up with engineered static Hamiltonians. Instead, these FBS's exhibit a unique DTC feature. Specifically, each scar quasienergy may be shifted considerably under perturbation. However, the interplay of strong interactions and drivings equalizes the scar level shifts, which is proved to all orders in our perturbative treatment. Then, the quasienergy difference $ \omega_0 $ between FBSs remains invariant and enforces the persisting $ 2\pi/\omega_0 $-periodic DTC. Rigid scar level spacing here resembles the ``spectral pairing rigidity" for all Floquet eigenstates in MBL DTCs~\cite{Khemani2016,Keyserlingk2016}. Also, such a mechanism allows for rather generic perturbations compared with preexisting scar models typically relying on microscopic details to achieve configuration separations~\cite{Bernien2017,Bluvstein2021,Maskara2021,Sugiura2021,Mizuta2020,Choi2019,Zhao2020,Mukherjee2020,Turner2018,Khemani2019a,Desaules2021,Scherg2021,Su2022}.
	Thus, our analytical solutions not only offer a more definitive understanding of clean DTC mechanisms, but also point out a new way of constructing scars showing peculiar spectral orders characteristic of Floquet systems.

\begin{figure}
		[h]
		\parbox{3.8cm}{
			\includegraphics[width=3.5cm]{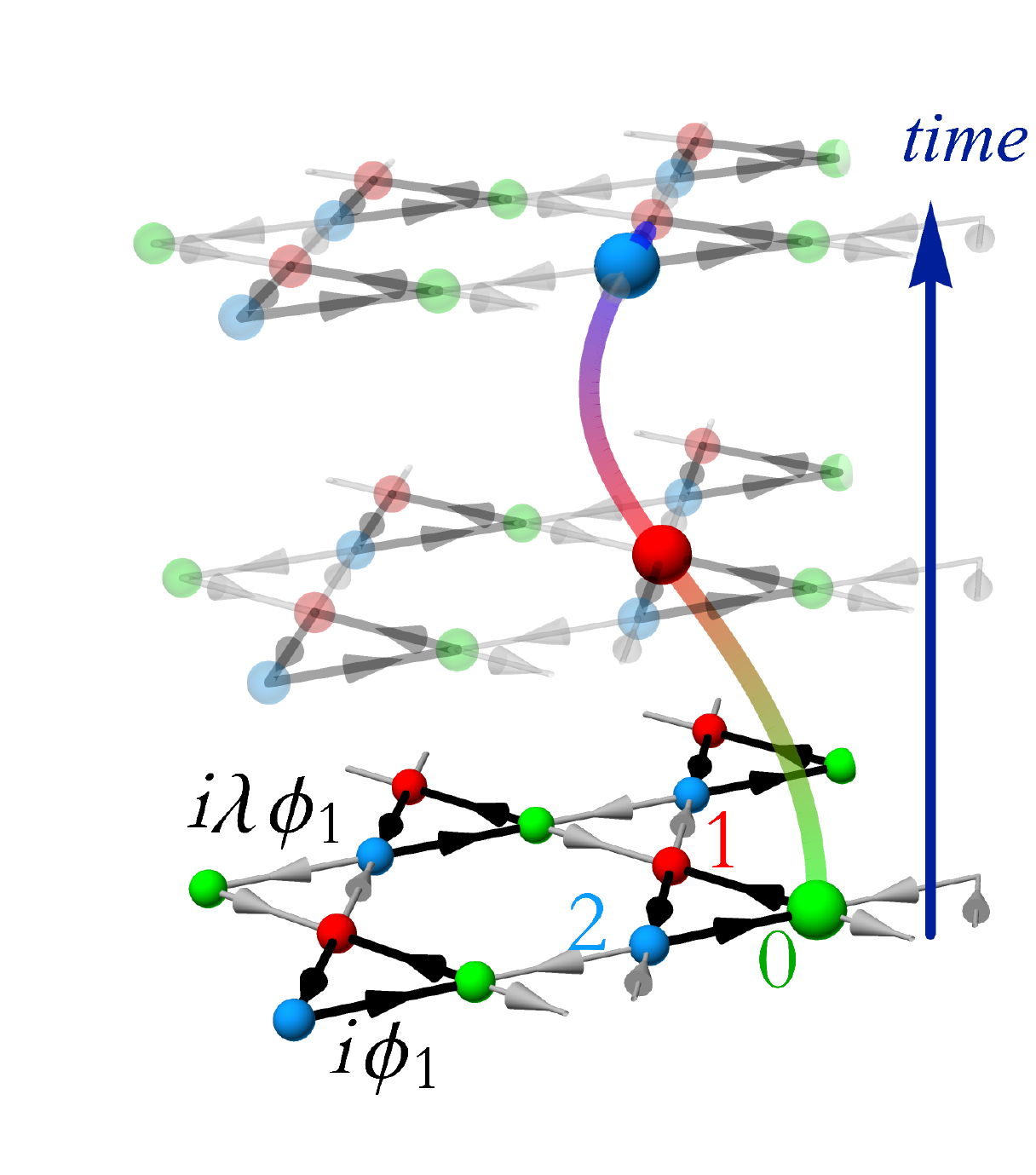} \\ (a) Lattice and phenomena\\ \quad \\
			\includegraphics[width=3.8 cm]{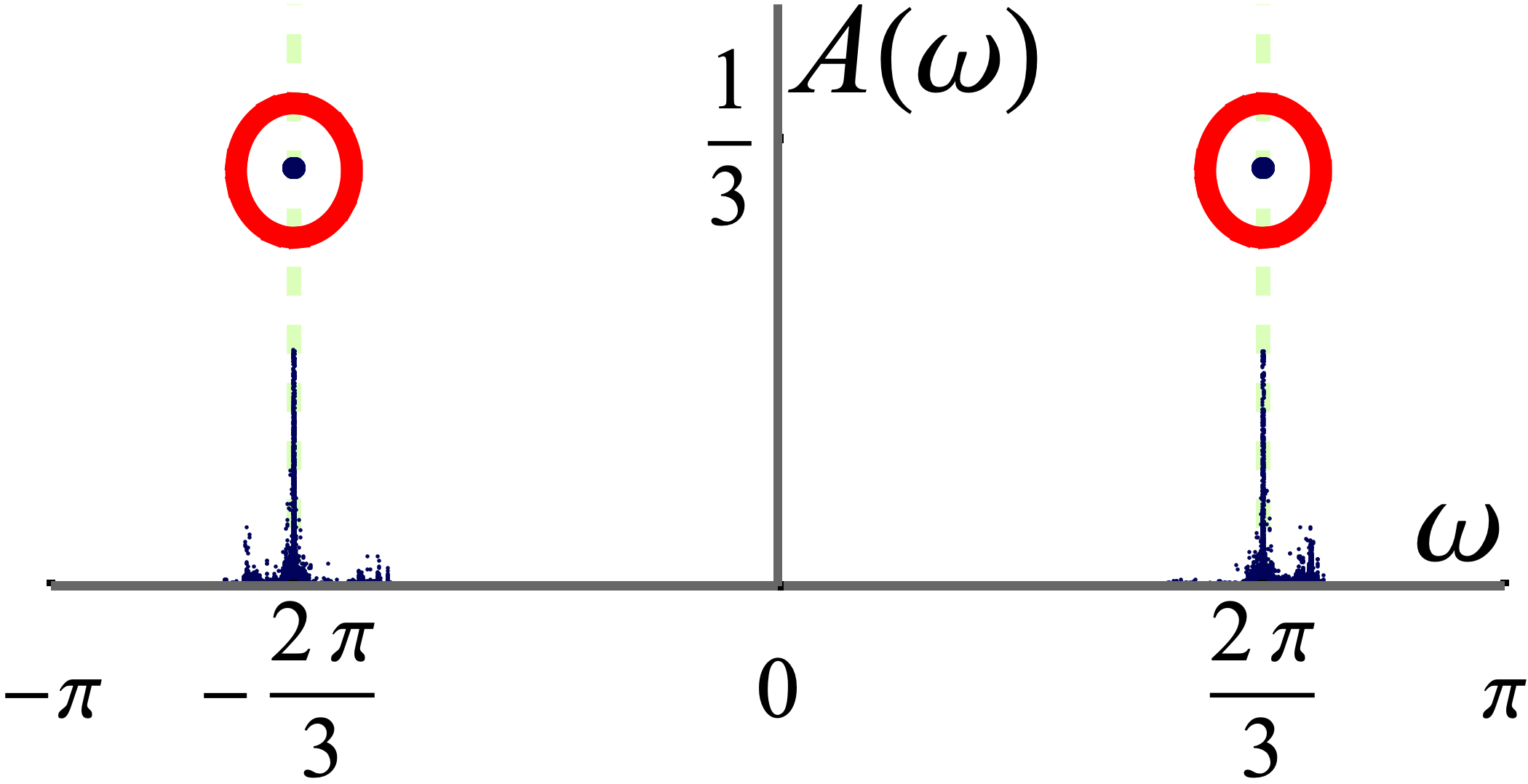}\\
			(c) Eigenstate correlations
		}
		\parbox{4.6cm}{
			\includegraphics[width=4.5cm]{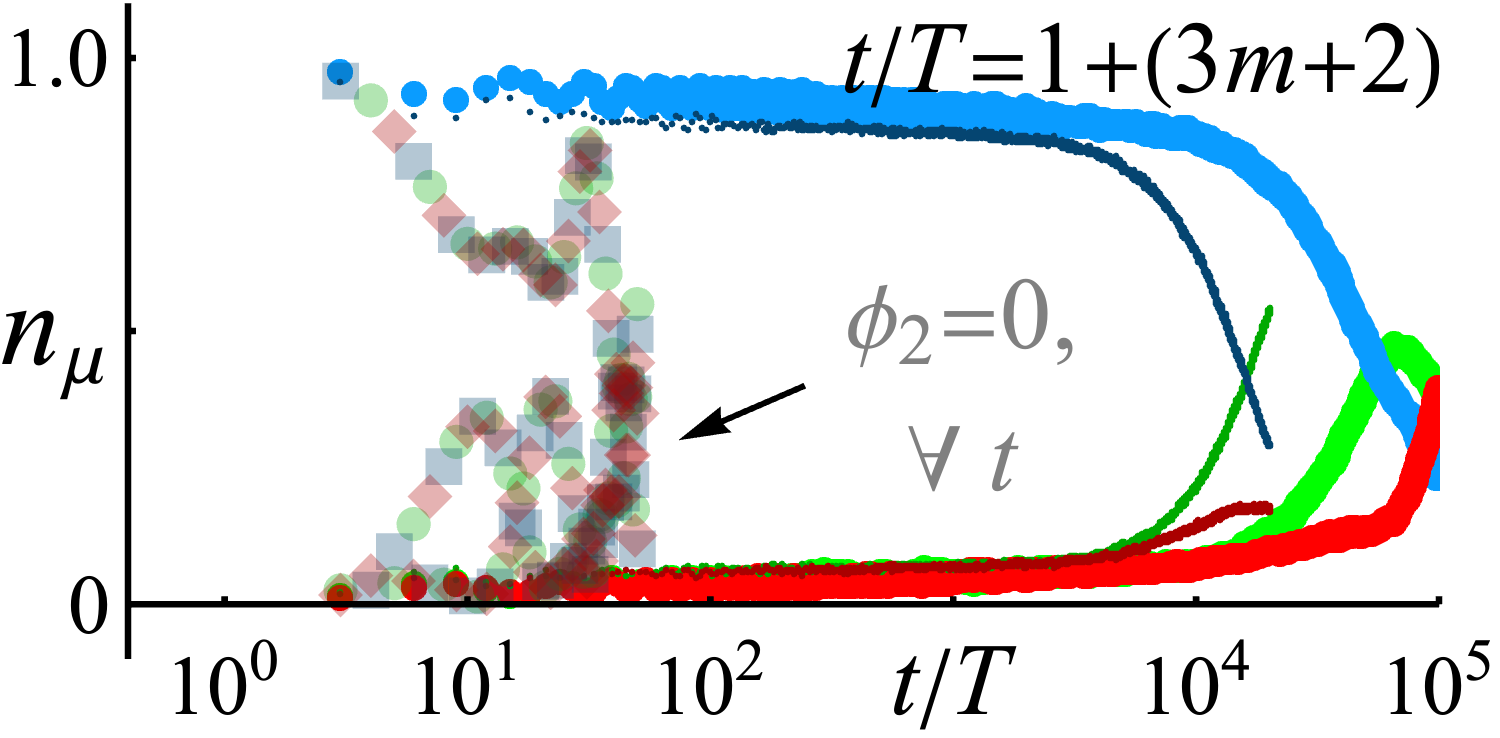}
			\\
			\includegraphics[width=4.5cm]{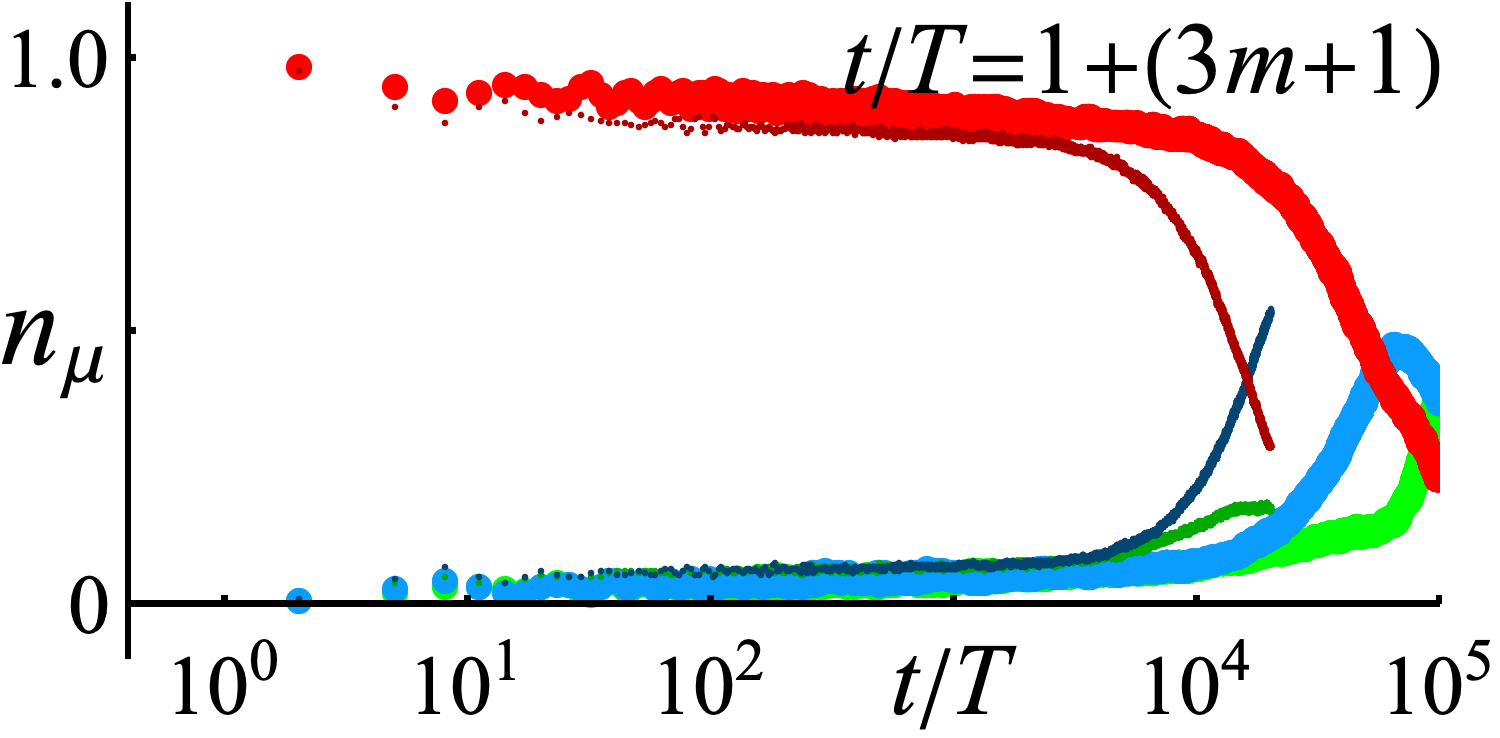}
			\\
			\includegraphics[width=4.5cm]{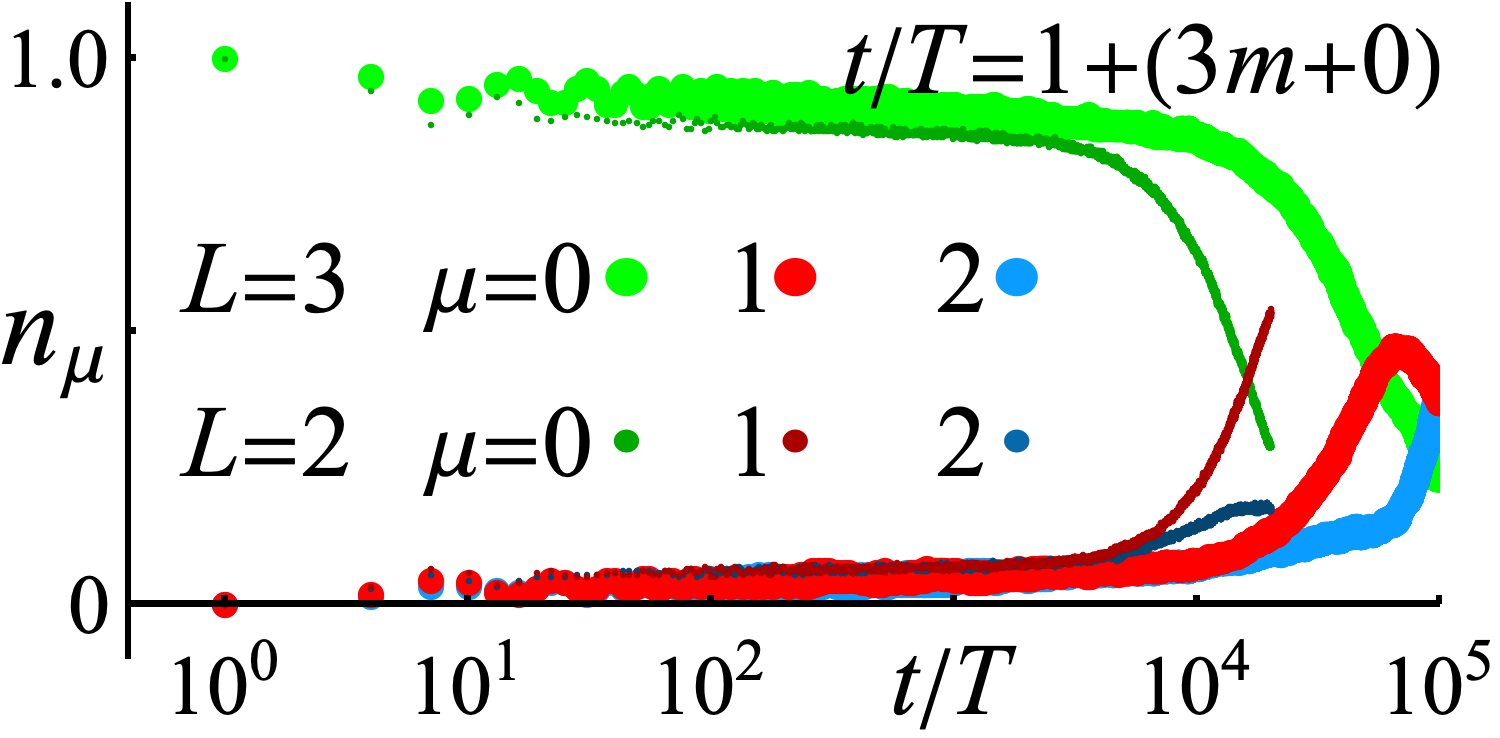}\\
			(b) Dynamics ($ m\in\mathbb{Z} $)
		}
		\caption{\label{fig:model}
			(a) Trimerized kagome lattice and the schematic illustration of DTC dynamics. Triangles with strong/weak bonds are denoted by black/gray colors, with $ +i $ hopping directions indicated by arrows.
			(b) Particle dynamics $ n_\mu(NT) = \sum_{\boldsymbol{r}}\langle \psi_{\text{ini}} | (\hat{U}_F^\dagger)^N n_{\boldsymbol{r}\mu} \hat{U}_F^N |\psi_{\text{ini}} \rangle $, with the initial state  $ |\psi_{\text{ini}}\rangle $ at $ t/T=1 $ that all particles locate at a single site $ \boldsymbol{r}=\boldsymbol{0}, \mu=0 $. To facilitate reading, data is grouped into 3 sets at $ t\mod 3T = 0, 1, 2 $ respectively. For comparison, the non-interacting $ \phi_2=0, L=3 $ case is  plotted at all $ t $ in the upper panel as translucent dots.
			(c) Temporal correlation functions indicating infinite-time response frequencies ($ L=3 $). Unless denoted otherwise, $ N_b = 5 $, $ \phi_1 = 2\pi/3\sqrt3$, $\lambda=0.1$, $\phi_2=1.1$, $\theta_{1,2,3} = (0.1,0.2,-0.3) $.
 		}
	\end{figure}

	{\noindent\em\color{blue}Model and phenomena} ---
	We consider bosons evolving under a Hamiltonian $\hat{H}(t+T) = \hat{H}(t)$ that is toggled  between two settings repetitively within each period $ T $: 
	\begin{align}\nonumber
		\hat{H}_1T/2\hbar &= \phi_1 \sum_{\boldsymbol{r}, \mu\ne\nu} 
		if_{\mu\nu}
		\left[  
		\hat{\psi}_{\boldsymbol{r}\mu}^\dagger \hat{\psi}_{\boldsymbol{r}\nu} 
		+ \lambda \hat{\psi}_{\boldsymbol{r}+\boldsymbol{e}_{\mu}, \mu }^\dagger 
		\hat{\psi}_{\boldsymbol{r}+\boldsymbol{e}_{\nu}, \nu}
		\right] , 
		\,
		t\in[0,T/2)
		\\
		\label{eq:model}
		\hat{H}_2T/2\hbar &=
		\sum_{\boldsymbol{r},\mu}
		\left[
		\phi_2 \hat{n}_{\boldsymbol{r}\mu} (\hat{n}_{\boldsymbol{r}\mu}-1)
		+ \theta_\mu \hat{n}_{\boldsymbol{r}\mu}
		\right], \, t\in[T/2,T).
	\end{align}
	Here, $\hat{H}_1$ describes the hopping of non-interacting bosons in a trimerized kagome lattice with complex hopping amplitudes, as shown in Fig.\ \ref{fig:model}(a), while $\hat{H}_2$ describes the combination of on-site single-particle and interaction energy shifts.
	Dimensionless parameters $ (\phi_1, \lambda, \phi_2, \theta_\mu) $ characterize the Floquet operator
	$ \hat{U}_F = P_t e^{- (i/\hbar) \int_0^T dt \hat{H}(t)} = e^{-i\hat{H}_2T/2\hbar} e^{-i\hat{H}_1T/2\hbar} $. $ \hat{\psi}_{\boldsymbol{r}\mu} $ and $ \hat{n}_{\boldsymbol{r}\mu} = \hat{\psi}_{\boldsymbol{r}\mu}^\dagger \hat{\psi}_{\boldsymbol{r}\mu} $ are
	annihilation and particle number operators respectively, for $ L^2 $ unit cells $ \boldsymbol{r} =m_1 \boldsymbol{e}_1 + m_2\boldsymbol{e}_2 $ ($ m_{1,2} = 0,1,\dots,L-1 $) and three sublattices $ \mu,\nu=0,1,2 $. 
	Here $ \boldsymbol{e}_{1,2} $ are Bravais vectors for kagome lattices and $ \boldsymbol{e}_0 \equiv \boldsymbol{0} $. $ if_{\mu\nu} = (1+2e^{2\pi i (\mu-\nu)/3})/\sqrt3 = \pm i $ specifies the $ +i $ hopping directions in Fig.~\ref{fig:model} (a). Note that $ \sum_\mu\theta_\mu = 0 $ can always be achieved by subtracting $ (N_b/3) \sum_\mu \theta_\mu $ from $ \hat{H}_2T/2\hbar $, where total bosons $ N_b = \sum_{\boldsymbol{r}\mu} \hat{n}_{\boldsymbol{r}\mu} $.

	DTC dynamics obtained by exact diagonalization is briefly shown in Fig.~\ref{fig:model}. When $ \lambda\rightarrow0 $, $ \hat{H}_1 $ enters the strongly trimerized regime composed of disconnected triangles, where $ \pi/2 $-fluxes equalize the spacing between single-particle flat bands $ \omega_n=0,\pm\sqrt3\phi_1 $ ($\hat{U}_F|\omega_n\rangle = e^{i\omega_n}|\omega_n\rangle$). Then,  $ \phi_1=2\pi/3\sqrt3 $ leads to $ 3T $ ballistic oscillations for particles $ \hat{U}_F^\dagger \hat{\psi}^\dagger_{\boldsymbol{r}, \mu=0,1,2} \hat{U}_F = \hat{\psi}_{\boldsymbol{r}, \mu=1,2,0} $  breaking the Hamiltonian time translation symmetry of $ T $, as in Fig.~\ref{fig:model} (a). Frequencies given by single-particle physics are, of course, unstable against perturbations. It is then the hallmark for DTC where strong interactions $ \phi_2 $ stabilize the $ 3T $ periodicity without fine-tuning, see Fig.~\ref{fig:model} (b). Late time dynamics can be further confirmed by the temporal correlation functions $ C(\omega)= \sum_{N=-\infty}^\infty \frac{e^{i\omega N }}{2\pi} \sum_n \langle \omega_n |\hat{P} (N) \hat{P} (0) |\omega_n \rangle = \sum_{mn} \delta(\omega-\omega_{mn}) A(\omega_{mn}) $ for the sublattice density bias, i.e. $ \hat{P}(N) = \left(\hat{U}_F^\dagger\right)^N N_b^{-1}\sum_{\boldsymbol{r}} (\hat{n}_{\boldsymbol{r}0} - \hat{n}_{\boldsymbol{r}1} ) \hat{U}_F^N $. 
	Note that the summation $ N $ is over {\em infinite} time without truncation. The spectral weight $ A(\omega_{mn}) = |\langle \omega_m |\hat{P} |\omega_n\rangle |^2 $, $ \omega_{mn} = \omega_m - \omega_n $ in Fig.~\ref{fig:model} (c) showing strong peaks at frequencies $ \omega_0 \rightarrow \pm2\pi/3 + O(1/D) $ verifies long-time oscillation periods $ 2\pi T/|\omega_0| \rightarrow 3T $. The small  deviation $ O(1/D) $ suppressed by Hilbert space dimension $ D $ gives an envelop modulation in Fig.~\ref{fig:model} (b) as noticed previously for both MBL~\cite{Else2016,Keyserlingk2016} and clean~\cite{Huang2018} DTCs.

	The above phenomena may be viewed from several angles. Particularly, in the case of complete trimerization, $\lambda = 0$, the two-dimensional lattice breaks up into isolated trimers. DTCs observed in this case is then explained simply as that of a microscopic three-site chiral system similar to Ref.~\cite{Pizzi2019}. If we were to regard intertrimer coupling as simply opening up each one-trimer DTC to an external bath composed of other trimers, we might expect the overall DTC dynamics to be destroyed over short time at $\lambda \neq 0$ \cite{Lazarides2017}.  Yet, such expectations contradict results in Fig.~\ref{fig:model} (b) (c).  Below, we offer an explanation that DTCs in the coupled-trimer regime is stabilized by a special class of scar Floquet eigenstates each spanning over the entire two-dimensional lattice.

	\begin{figure}
		[h]
		\parbox[b]{8.6cm}{
			\parbox[b]{6.3cm}{
				\parbox{6.2cm}{
					\parbox[b]{3.4cm}{	\includegraphics[width=3.4cm]{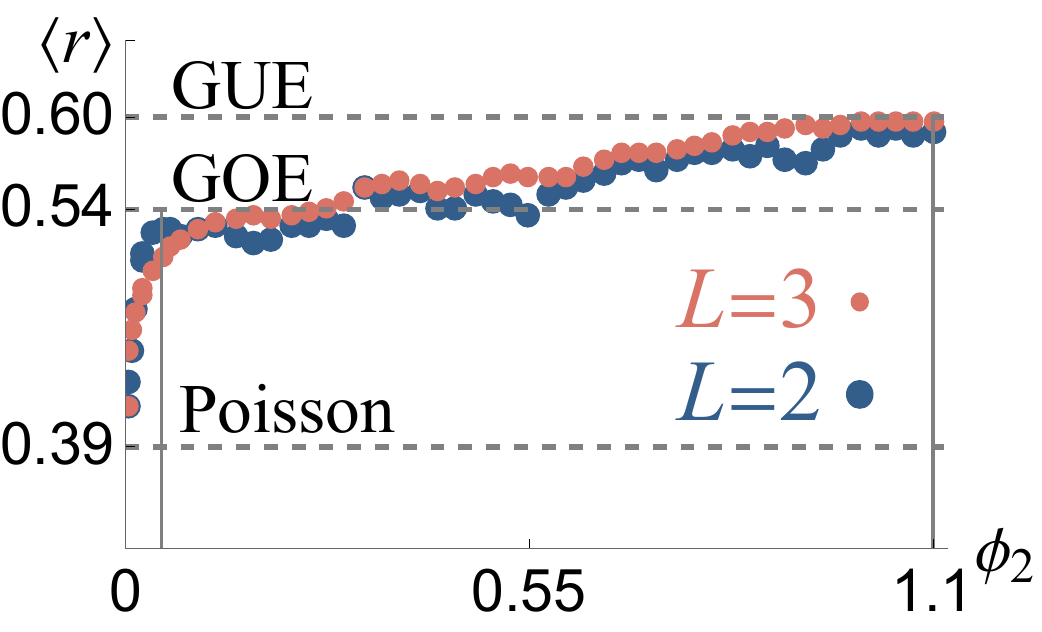}
						\\ (a) Level statistics }
					\quad
					\parbox[b]{2cm}{
						\raggedright
						\includegraphics[width=1.65cm]{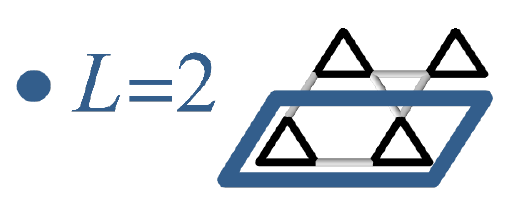}
						\includegraphics[width=2cm]{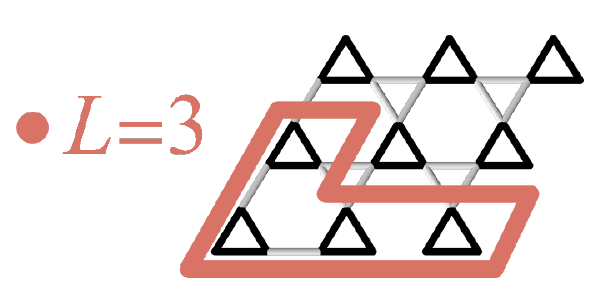}\\
						(b) EE cut
					}
				}\\
				\parbox[b]{6.4cm}{
					\parbox[b]{3.15cm}{
						\includegraphics[width=3.2cm]{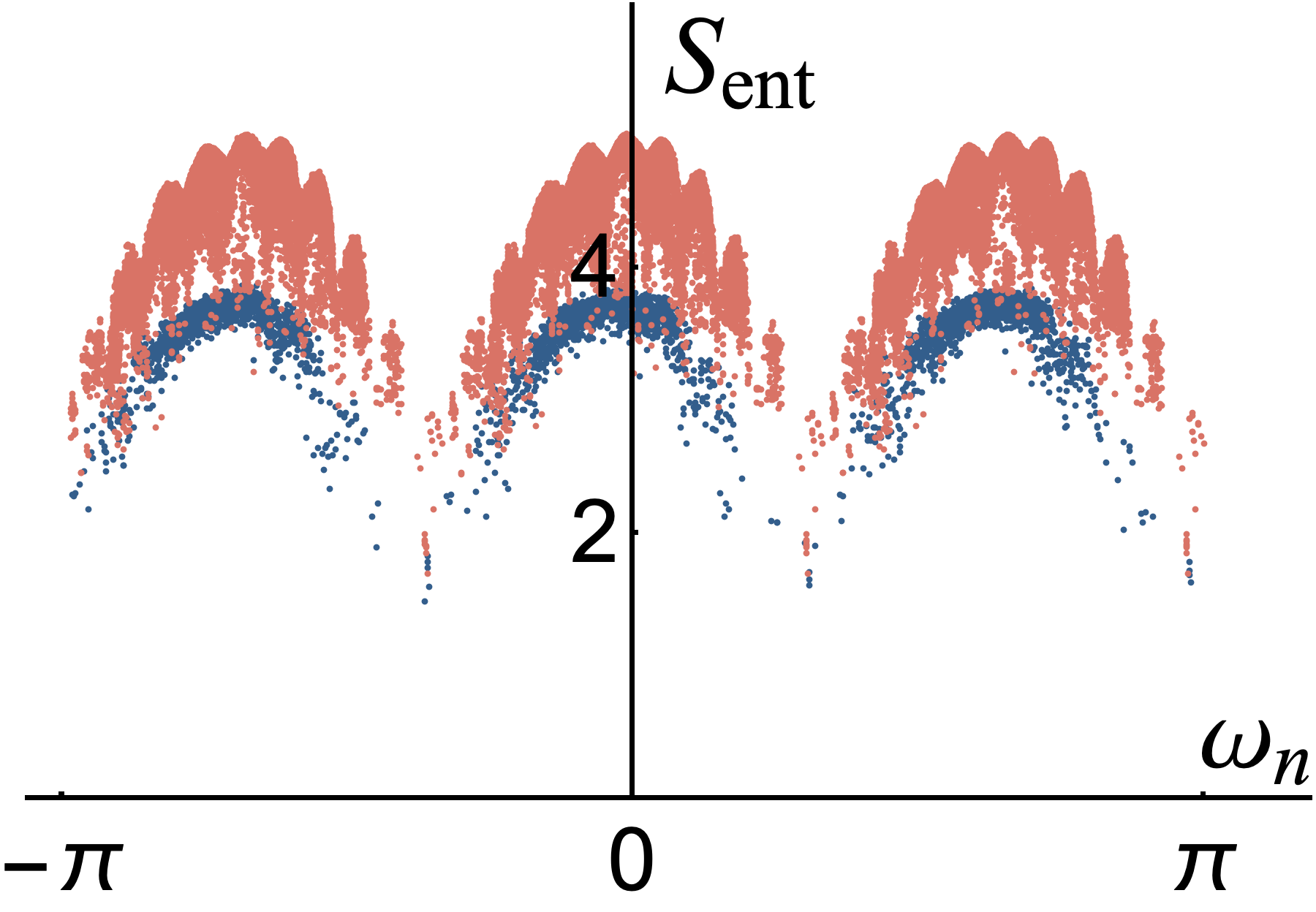}
						\\ (c) $ \phi_2=0.05 $ }
					\parbox[b]{3.1cm}{
						\includegraphics[width=3.2cm,trim=0 0.2cm 0 0]{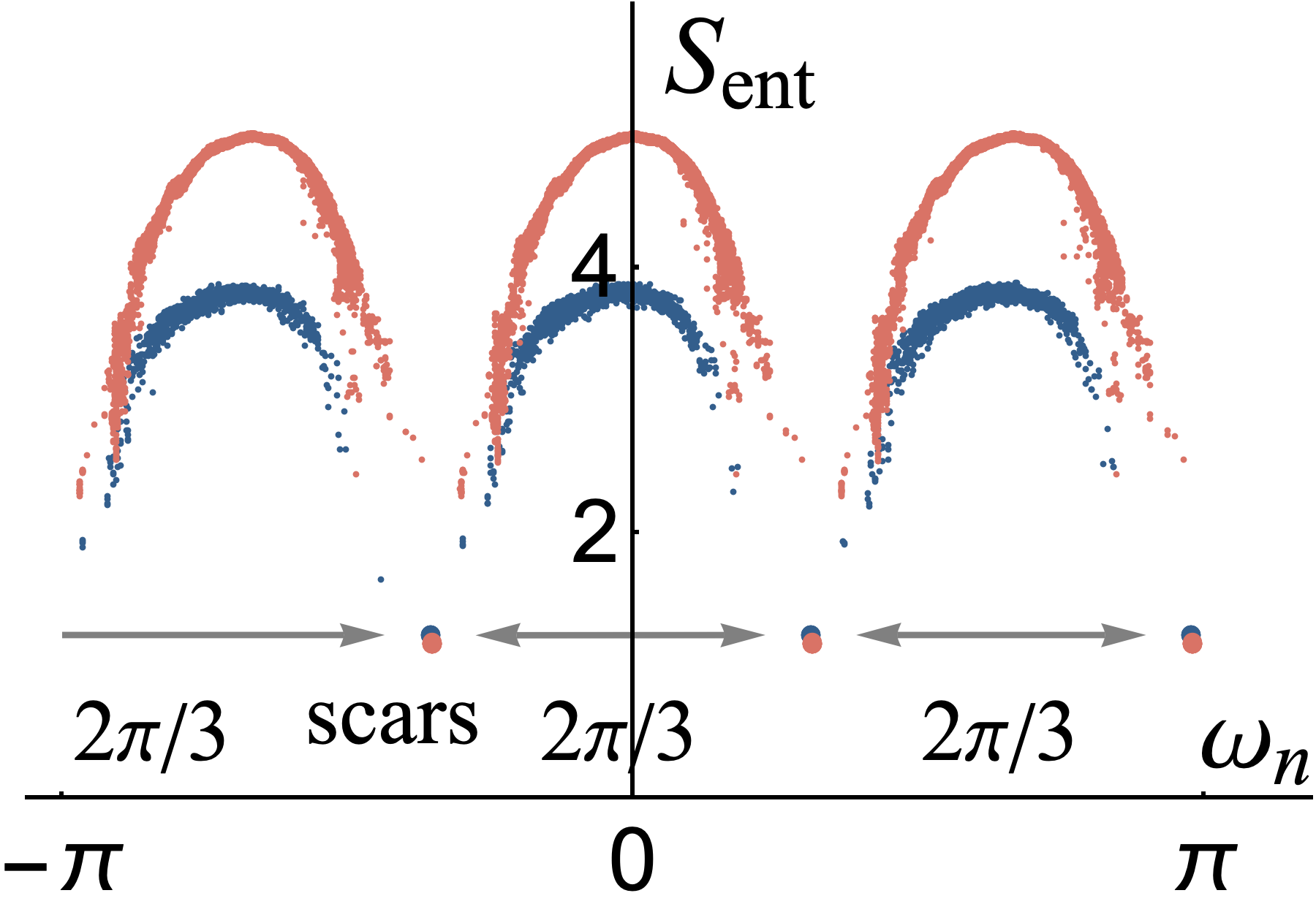}
						\\ (d) $ \phi_2=1.1 $ }
				}
			}
			\parbox[b]{2.2cm}{
				\includegraphics[width=2.3cm]{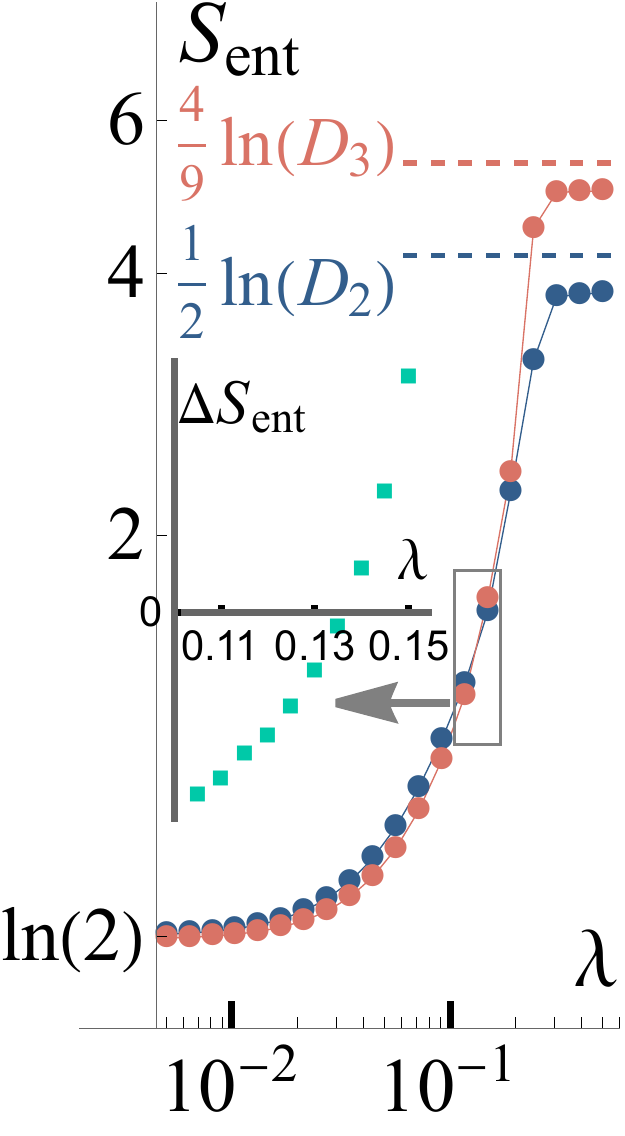}
				\\
				(e) Lowest $ S_{\text{ent}} $ for $ \phi_2=1.1 $
			}
		}
		\caption{\label{fig:r_ee2}
			(a) $ \langle r\rangle $ shows generic ergodicity.
			(b) Subsystem for computing EE. 
			(c) -- (d) Eigenstate EE in (c) proximate-integrable and (d) DTC regimes, where low-entropy scars in (d) are highlighted by larger dots. 
			(e) Lowest EE approaching size-insensitive values at small $ \lambda $ and volume law $ S_{\text{ent}} \sim (N_{\text{sub}}/3L^2)\ln(D_L) $ at large $ \lambda $. Here $ D_L $ is the total Hilbert space dimension and $ N_{\text{sub}} $ the subsystem site number enclosed in (b). Inset shows $ \Delta S_{\text{ent}} = S_{\text{ent}}^{(L=3)} - S_{\text{ent}}^{(L=2)} $ near the crossing $ \lambda_0\approx 0.135 $.
			Unless specified otherwise, in all plots parameters are the same as in Fig.~\ref{fig:model}. Blue (or red) colors denote $L=2$ (or $L=3$) respectively.
		}
	\end{figure}

	{\noindent\em\color{blue}Identifying scars ---} Quantum scars are rare non-ergodic eigenstates within an eigenstructure that is otherwise thermalizing~\cite{Serbyn2021}.  Numerical calculations confirm the overall thermalizing, non-integrable nature of our model system.  Specifically, we point to two signatures of non-integrability: level-spacing statistics and entanglement entropy.
	
	Consider first the level spacing. Ordering quasi-energies as $\omega_{n} < \omega_{n+1}$, following Ref.\ \cite{Atas2013}, we test for ergodicity by calculating the level spacing ratios  $ r_n=\min(\delta_n, \delta_{n+1}) / \max(\delta_n, \delta_{n+1}) $ for consecutive gaps $ \delta_{n}=\omega_{n+1}-\omega_n $.  
	Clearly from Fig.~\ref{fig:r_ee2} (a), except for a vanishingly small region in proximity to single particle limit $ \phi_2\rightarrow0 $, our model is generically far from the integrable Poissonian case $ \langle r \rangle \rightarrow 0.39$. We also note a crossover between two ergodic Gaussian orthogonal/unitary ensembles (GOE/GUE) purely by different drivings, an interesting feature previous seen in spin models~\cite{Regnault2016}.

	
	We next exploit the entanglement entropy (EE) to examine each Floquet eigenstate $ |\omega_n\rangle $. Reduced density matrices $ \rho_{A} = \text{Tr}_B(|\omega_n\rangle \langle \omega_n| ) $ for subsystem $ A $ (region enclosed by highlighted paths in Fig.~\ref{fig:r_ee2} (b)) can be formed  by tracing out the remaining part $ B $ in real space. The EE $ S_{\text{ent}} = -\text{Tr} \left( \rho_A \ln \rho_A \right) $ then shows that in both proximate-integrable (Fig.~\ref{fig:r_ee2} (c)) and DTC (Fig.~\ref{fig:r_ee2} (d)) regimes, majority eigenstates do exhibit the typical arch shape for $ S_{\text{ent}} $ whose values increase with Hilbert space dimensions~\cite{DAlessio2016}.
	The narrow distribution of EE for eigenstates of similar quasi-energy in the DTC regime  confirms that majority arch eigenstates are ergodic~\cite{DAlessio2016}, in consistent with $ \langle r\rangle $ results previously.

	However, in the DTC regime, additional  non-ergodic states are observed.  As exhibited in Fig.\ \ref{fig:r_ee2}(d), we identify precisely $ 3L^2 $  low $ S_{\text{ent}} $ scar states (each scar dot in the figure is  $ L^2 $-fold degenerate). Each set of scars separates from the others by quasienergy $ |\Delta E|\rightarrow 2\pi/3 $, corresponding to exactly the DTC frequency in Fig.~\ref{fig:model} (c). The scaling of lowest EE  in Fig.~\ref{fig:r_ee2} (e) shows a system size $ L $ insensitive scar EE for $ \lambda\rightarrow0 $. 
	With increasing $ \lambda $, a possible transition is observed around $ \lambda_0\approx 0.135 $~\footnote{Due to limited sizes accessible here, we would postpone a more comprehensive examination of criticality to future work and only take $ \lambda_0 $ as a reference scar vanishing point.}, after which all eigenstates approach the volume law 
	ergodic limit.

	We have confirmed numerically that parameters in Fig.~\ref{fig:r_ee2} (c) give rather short DTC lifetime, unlike the lifetime shown in Fig.~\ref{fig:model} (b) for parameters in Fig.~\ref{fig:r_ee2} (d). It strongly indicates that the DTC behaviors here are intimately associated with scars rather than (approximate) overall integrability.


	{\noindent\em\color{blue}Analytical results for FBS ---} To characterize these quantum scars further, we work in the many-body momentum basis~\cite{Sandvik2010} 
	$ |\boldsymbol{k},\{n_{\boldsymbol{r},\mu}\} \rangle =
	(1/L)\sum_{m_1, m_2 =0}^{L-1}
	e^{(2\pi i/L) (k_1 m_1 + k_2 m_2)}
	\left| \{n_{\boldsymbol{r}+m_1\boldsymbol{e}_1 + m_2\boldsymbol{e}_2,\mu}\} \right\rangle  $  constructed from Fock basis $ | \{n_{\boldsymbol{r}\mu}\}\rangle = \prod_{\boldsymbol{r}\mu} 
	\left[
	\left(\hat{\psi}_{\boldsymbol{r}\mu}^\dagger\right)^{n_{\boldsymbol{r}\mu}}
	/
	\sqrt{n_{\boldsymbol{r}\mu}!}
	\right] |0\rangle   $.
	Here $ \{n_{\boldsymbol{r}\mu}\} $ specifies occupation numbers at different sites, and $ \boldsymbol{k} \sim  k_{1,2} = 0,1,\ldots,L-1 $. Then, translation-invariant  $ \hat{U}_F $ are block-diagonalized 
	$ \langle \boldsymbol{k},\{n_{\boldsymbol{r}\mu}\}|U_F| \boldsymbol{k}', \{n_{\boldsymbol{r}\mu}'\}\rangle \sim \delta_{\boldsymbol{k},\boldsymbol{k}'} $. 
	Each $ \boldsymbol{k} $ sector would be shown later to host $ 3 $ scar states, leading to the $ 3\times L^2 $-fold scars in Fig.~\ref{fig:r_ee2} (d).

	It is helpful to write down the solution $ U_F |\boldsymbol{k},\ell,\{n_{\boldsymbol{r}\mu} \} \rangle = e^{iE(\ell,\{n_{\boldsymbol{r}\mu}\})} |\boldsymbol{k},\ell,\{n_{\boldsymbol{r}}\} \rangle $ to Eqs.~\eqref{eq:model} at the anchor point $ \lambda=0 $,
	\begin{align}\label{eq:lnk}
		\left| \boldsymbol{k},  \ell, \{n_{\boldsymbol{r}\mu}\} \right\rangle 
		&=\,
		\frac{1}{\sqrt3} \sum_{m=0,1,2} e^{-i(\frac{2\pi m}{3}\ell -\alpha_m)} 
		|\boldsymbol{k}, \{n_{\boldsymbol{r},\mu+m \text{ mod } 3} \} \rangle,\\
		\label{eq:energy}
		E\left(\ell, \{n_{\boldsymbol{r}\mu}\}\right) &=  \frac{2\pi}{3} \ell  + \phi_2 \sum_{\boldsymbol{r}\mu} n_{\boldsymbol{r} \mu} (n_{\boldsymbol{r}\mu} - 1),  \qquad
		\ell = 0, \pm 1,
	\end{align}
	where	$ \alpha_0 = 0$,
	$
	\alpha_{1} = \sum_{\boldsymbol{r}\mu}\theta_\mu n_{\boldsymbol{r}\mu} 
	$,
	and $	\alpha_{2} = -\sum_{\boldsymbol{r}\mu} \theta_\mu n_{\boldsymbol{r}, \mu + 2 \text{ mod } 3}  $. 
	Each eigenstate populates 3 sublattices $ \mu=0,1,2 $ coherently, and therefore an arbitrary Fock state $ |\{n_{\boldsymbol{r}\mu}\} \rangle $, usually taken as initial states, will simultaneously overlap with all three branches $ \ell=0,\pm1 $ separating from each other by quasi-energy $ |\Delta E| = 2\pi/3 $. Then, observables diagonal in the Fock basis, such as $ \hat{O} = \hat{n}_{\boldsymbol{r}\mu} $ or $ \hat{O} =  \hat{P} = N_b^{-1} \sum_{\boldsymbol{r}} (\hat{n}_{\boldsymbol{r}0} - \hat{n}_{\boldsymbol{r}1}) $, will demonstrate an oscillation $ \langle \hat{O}\rangle (t) \sim  c_1^* c_2 \langle \boldsymbol{k}_1, \ell_1, \{n_{\boldsymbol{r}\mu}\} | \hat{O} |\boldsymbol{k}_2, \ell_2, \{n_{\boldsymbol{r}\mu}\}  \rangle e^{-i\Delta E t/T} + c.c. $ with periodicity $ 2\pi T/\Delta E = 3T $.
	
	Spectral pairing $ \Delta E $ for majority eigenstates in Eqs.~\eqref{eq:lnk} \eqref{eq:energy} is, as expected, unstable against perturbations. The crucial difference here from the disordered case~\cite{Khemani2016,Keyserlingk2016,Else2016,Yao2017} is the uniform  interaction strength $ \phi_2 $ in Eq.~\eqref{eq:energy}, which results in an enormous Floquet emergent degeneracy. Specifically, consider the combination $ Q_a = \{ (q^{(a)}_j, N^{(a)}_j) |j=1,2,\ldots,M\} $ for, i.e. $ q^{(a)}_j $ copies of sites each hosting $ N^{(a)}_j\ge0 $ particles. In terms of the Hubbard interaction $ \phi_2 \sum_{\boldsymbol{r}\mu} n_{\boldsymbol{r}\mu} (n_{\boldsymbol{r}\mu}-1) = \phi_2 \sum_j q^{(a)}_j N^{(a)}_j(N^{(a)}_j-1) $, each $ Q_a $ manifold contains degenerate levels of different $ \{n_{\boldsymbol{r}\mu}\} $ as $ \text{deg}(Q_a) = (3L^2)!/\prod_{j=1}^M q^{(a)}_j! $. The degeneracy, though partially lifted by $ 2\pi\ell/3 $ in Eq.~\eqref{eq:energy}, leads to the instability that a small perturbation could generally trigger a reconstruction for extensive numbers of eigenstates in Eq.~\eqref{eq:lnk} with different configurations $ \{n_{\boldsymbol{r}\mu}\} $, leading to the ergodicity as indicated by Fig.~\ref{fig:r_ee2}. Correspondingly, a Fock initial state would overlap with large numbers of eigenstates with different quasienergies without rigid spectral pairings.

	To identify FBS, we then seek for manifolds with low degeneracy. Except for a homogeneous distribution $ n_{\boldsymbol{r}\mu} = N_b/3L^2 $  (deg=$ 1 $) without dynamical signatures, the lowest degenerate $ \{n_{\boldsymbol{r}\mu}\} $ deposit all $ N_b $ bosons into a single site $ n_{\boldsymbol{r}\mu} = \delta_{\boldsymbol{r},\boldsymbol{r}_0} \delta_{\mu,\mu_0} N_b $. There are apparently $ 3L^2 $ such $ \{n_{\boldsymbol{r}\mu}\} $ with $ N_b $ bosons allocated into different sites $ (\boldsymbol{r}_0, \mu_0) $. They compose the FBS eigenstates
	\begin{align}\label{eq:scar}
		|\boldsymbol{k}, \ell, N_b \rangle = \frac{1}{\sqrt{3}} \sum_{m=0,1,2} 
		e^{-i(\frac{2\pi m}{3}\ell - \alpha_m)} 
		|\boldsymbol{k}, \{ n_{\boldsymbol{r}\mu} =  \delta_{\boldsymbol{r},\boldsymbol{0}} \delta_{\mu,m}N_b \} \rangle,
	\end{align}
	with quasienergy $ E_{\text{scar}}(\ell) = 2\pi \ell/3 + \phi_2 N_b(N_b-1) $.
	The $ 3L^2 $ FBSs equally partition into $ L^2 $ conserved many-body momentum $ \boldsymbol{k} $ sectors, each hosting 3 scars with $ \ell = 0, \pm1 $. Spatial translation symmetry then forbids hybridizing eigenstates of different $ \boldsymbol{k} $, and temporal translation symmetry protects the conserved quasienergy  separating different $ \ell $ by $ |\Delta E| = |E_{\text{scar}}(\ell+1) - E_{\text{scar}}(\ell)| =2\pi/3 $. Therefore, FBS's experience no degenerate-level perturbations.
	
	It still remains to consider non-degenerate perturbations. In particular, the periodicity $ 2\pi $ of Floquet quasienergy constrains Hubbard-interaction gap for different $ Q_a $ to be of the order unity. Then, one may expect each scar level to receive an energy correction $ \sim \lambda^2 $ (of Fermi golden rule type), resulting in fast detuning within $ t\sim T/\lambda^2 \sim 100T $ for $ \lambda=0.1 $. However, such estimations directly contradict Fig.~\ref{fig:model} (b). 
	
	The resolution turns out to be that all three $ \ell=0,\pm1 $ scars are shifted identically, such that their quasienergy difference, dubbed {\em spectral pairing} gap $ |\Delta E| = 2\pi/3 $~\footnote{In strongly disordered cases~\cite{Khemani2016,Keyserlingk2016,Else2016}, spectral pairing happens for all eigenstates due to many-body localization. Here we use the terminology to describe similar behaviors for scars but due to different reasons.}, is unchanged.
	In Supplemental Materials (SM)~\cite{supp}, we construct the strong-drive perturbation theory. For conciseness, we illuminate the essential physics below by elaborating results up to the second order in the perturbation series, while  higher orders cases are left to SM~\cite{supp}. 
	
	Arrange a Floquet operator in the form $ U_F = U_0 U' $, where $ U_0 $ corresponds to Eq.~\eqref{eq:model} at $ \lambda=0 $, and perturbations are factored into $ U' \equiv e^{i\lambda H'} $.  For our purposes, it is more than enough to take $ H' $ as a generic hopping Hamiltonian 
	$ H'=\sum J_{\boldsymbol{r}\mu\ne\boldsymbol{r}'\mu'} \hat{\psi}_{\boldsymbol{r}\mu}^\dagger \hat{\psi}_{\boldsymbol{r}'\mu'} $.  (See SM~\cite{supp} for factorization process).
	Scar quasienergy corrections 
	$ e^{i\tilde{E}_\text{scar}(\ell)} = e^{i(E_{\text{scar}}(\ell) + \sum_{\alpha=1}^\infty \lambda^\alpha E_\ell^{(\alpha)})} $ 
	up to the second order read 
	$ E^{(1)}_\ell = \langle \boldsymbol{k},\ell, N_b| H' |\boldsymbol{k}, \ell, N_b\rangle  $, 
	$ E^{(2)}_\ell = -\frac{1}{2}\sum_{(\ell',\{n_{\boldsymbol{r}\mu}\})}'
	|\langle \boldsymbol{k}, \ell', \{n_{\boldsymbol{r}\mu}\} | H' | \boldsymbol{k}, \ell, N_b\rangle |^2 \cot \frac{E_{\text{scar}}(\ell) - E(\ell',\{n_{\boldsymbol{r}\mu}\})}{2} $, 
	where summation $ \sum' $ excludes the scar eigenstate in consideration.
	Here, $ E^{(1)}_\ell = 0 $ is trivially identical for all $ \ell $. Importantly, Eqs.~\eqref{eq:lnk}--\eqref{eq:scar}   show that each term for $ E^{(2)}_\ell $ depends only on the difference $ (\ell - \ell') $. Due to $ 2\pi $ quasienergy periodicity, quantum numbers $ \ell $ in Eqs.~\eqref{eq:lnk} and \eqref{eq:energy} are only defined modulo $ 3 $. That allows for shifting dummy indices $ \ell' $ in the summation 
	$ 
	E^{(2)}_{\ell_1} \equiv \sum'_{\ell',\{n_{\boldsymbol{r}\mu}\}} \varepsilon(\ell_1 - \ell') 
	=
	\sum'_{\ell',\{n_{\boldsymbol{r}\mu}\}} \varepsilon (\ell_2 - (\ell'-\ell_1+\ell_2))
	=
	\sum'_{\ell'',\{n_{\boldsymbol{r}\mu}\}} \varepsilon(\ell_2-\ell'') = E^{(2)}_{\ell_2}
	$,
	proving the equality of energy corrections for all scars. SM~\cite{supp} also numerically verifies spectral pairing rigidity for Eq.~\eqref{eq:model} and against more generic bilinear perturbations.
	
	Importantly, it is exactly the Floquet spectrum periodicity that allows for shifting all three $ \ell $'s in Eq.~\eqref{eq:energy} by the same integer and end up with an identical set of levels, which is crucial for the above proof. In SM~\cite{supp}, we prove that the spectral pairing rigidity persists to {\em all perturbation orders} for FBS's. Therefore, $ O(L^2) $ initial states overlapping with multiple FBS's  separating by a rigid $ \omega_0 = 2\pi/3T $ will exhibit persisting $ 2\pi/\omega_0 = 3T $ DTC oscillations.

	Analytical identification of FBS's and proof for their spectral pairing rigidity are the main results of our work. They rely on three pivotal factors. First, {\em strong interactions} validate the starting point from Eqs.~\eqref{eq:lnk} and \eqref{eq:energy} for kicked Fock states. Second, {\em strong Floquet drivings} produce three identical $ \ell=0,\pm1 $ spectral plethora at $ \lambda=0 $, and the $ 2\pi $ quasienergy periodicity intrinsic of Floquet nature enables the rigid spectral pairing for FBS against perturbations. Third, {\em spatiotemporal translation symmetry} prevents FBS from mutual hybridization. Therefore, FBS's describe genuine strongly interacting Floquet matters in clean systems.
	
	{\em\noindent\color{blue} Numerical verification} ---
	Revisiting previous numerics can now be illuminating. Spectral function peaks in Fig.~\ref{fig:model} (c) derive from pairs of FBS's in Eq.~\eqref{eq:scar}, $ A(\omega_0)|_{\lambda\rightarrow0} = |\langle \boldsymbol{k}, \ell_1, N_b|\hat{P} |\boldsymbol{k}, \ell_1\pm1, N_b\rangle|^2 =1/3, |\omega_0| = 2\pi/3  $. The spectral pairing rigidity then stabilizes $ |\omega_0| $ against perturbation up to finite size effects, resulting in DTC oscillations in Fig.~\ref{fig:model} (b). Also, Eq.~\eqref{eq:scar} prescribes an $ L $-independent EE for FBS at $ \lambda\rightarrow0 $ (see SM~\cite{supp} for analytical calculation) as in Fig.~\ref{fig:r_ee2} (e).

	\begin{figure}
		[h]
		\parbox[b]{5.7cm}{
			\parbox[b]{3cm}{\includegraphics[width=2.5cm]{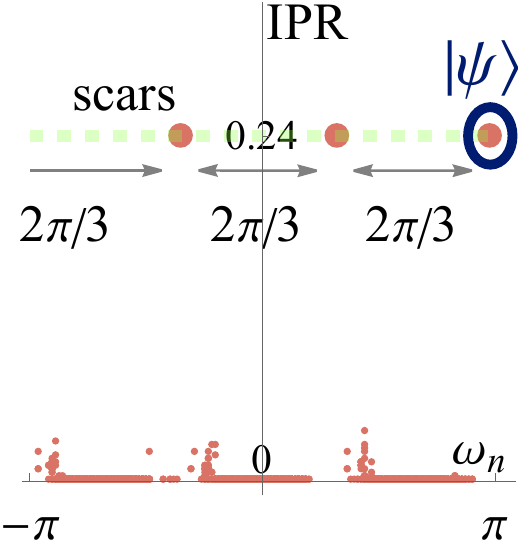}
				\\ (a) IPR for eigenstates
			}
			\parbox[b]{2.5cm}{\includegraphics[width=2.5cm]{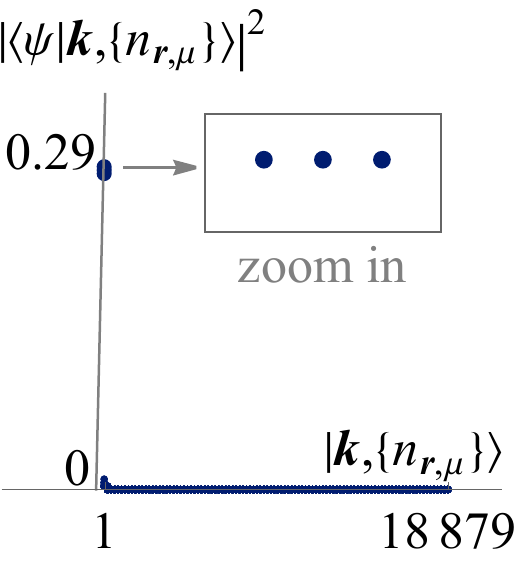}
				\\ (b) Component
			}\\
			\parbox[b]{5.7cm}{\includegraphics[width=5.7cm]{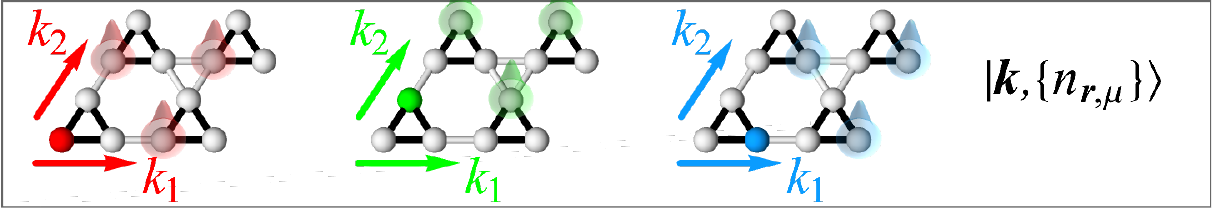} 
				\\(c) 3 dominant components in (b).
			}
			
		}
		\parbox[b]{2.7cm}{
			\includegraphics[width=2.1cm]{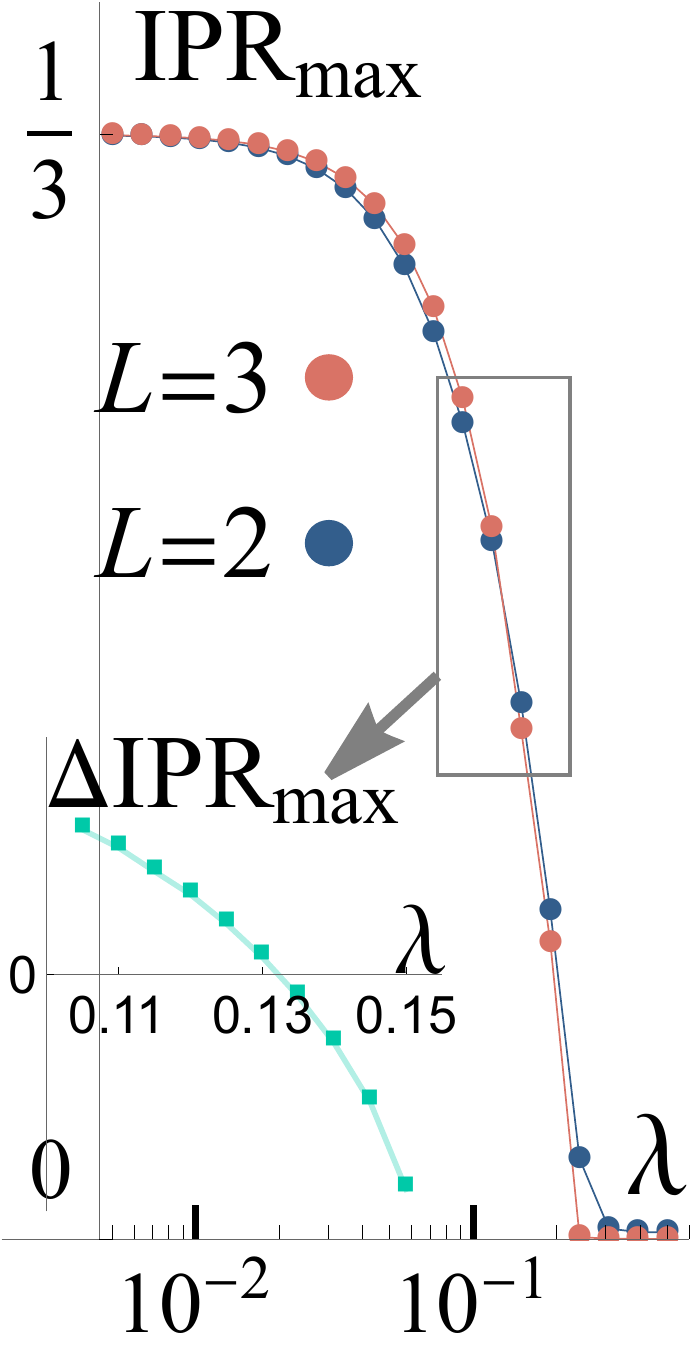}
			\\ (d) IPR$ _{\text{max}} $ scaling}
		\caption{\label{fig:eigV} 
			Structure of Floquet eigenstates in $ \boldsymbol{k} = \boldsymbol{0} $ sector. Other $ \boldsymbol{k} $ sectors show essentially the same results.
			{\bf (a)} Most eigenstates involve extensive number of basis $ |\boldsymbol{k},\{n_{\boldsymbol{r}\mu}\} \rangle $ leading to vanishing IPR, except for the 3 FBSs. 
			{\bf (b)} Expand for instance one FBS in the basis $ |\boldsymbol{k},\{n_{\boldsymbol{r}\mu}\}\rangle $, we see it is dominated by 3 components depicted in {\bf (c)}, exactly as given by Eq.~\eqref{eq:scar}.
			{\bf (d)} Scaling of the maximal IPR, where $\Delta \text{IPR}_{\text{max}} = \text{IPR}_{\text{max}}^{(L=3)} - \text{IPR}_{\text{max}}^{(L=2)} $ in the inset. 
			Parameters are the same as in Fig.~\ref{fig:model}, and $ L=3 $ for (a) (b).
		}
	\end{figure}

	Finally, we offer an efficient way to benchmark FBS by exploiting their peculiar $ \boldsymbol{k} $ space localization. A natural measure
	is then the momentum space inverse participation ratio IPR$ \,= \sum_{\{n_{\boldsymbol{r}}\} }|\langle \boldsymbol{k},\{n_{\boldsymbol{r}\mu}\} |\omega_n\rangle|^4 $, where scars would show exceptionally large IPR as in Fig.~\ref{fig:eigV} (a). Due to the absence of degenerate level hybridization, the original scar components in Eq.~\eqref{eq:scar} still dominate upon perturbation as in Fig.~\ref{fig:eigV} (b) and (c). The scaling of largest IPRs in Fig.~\ref{fig:eigV} (d) reproduces the reference transition $ \lambda_0\approx 0.135 $ as in Fig.~\ref{fig:r_ee2} (e).

	{\em\color{blue}\noindent Experimental relevance ---} 
	Small clusters studied above can be readily realized using the latest technology of quantum gas microscopes~\cite{Endres2016,Tai2017,Semeghini2021}, which allows for manipulation and detection with single-site resolutions. We now further discuss cases with finite filling fractions relevant to wider ranges of experiments. 
	
	In principle, previous analytical results show that initial states populating more than one unit cell will chiefly overlap with non-scar ergodic eigenstates. Therefore, a finite filling fraction among all unit cells will eventually lead to a thermalizing behavior without dynamical signatures. However, there could exist a finite and predictable time window before decay to observe the scar DTCs due to scar localization. 
	
	To show it, we first take a closer look at Fig.~\ref{fig:model} (b). The initial state of putting $ N_b $ bosons on one site overlaps with all FBS's (perturbed Eq.~\eqref{eq:scar}) in different $ (\boldsymbol{k},\ell) $ sectors; they interfere destructively everywhere except for the unit cell $ \boldsymbol{r} = \boldsymbol{0} $, resulting in a real-space localization. As such, two scar DTCs localized in different regions will take time to sense the presence of and affect each other $ \phi_2 n_{\text{ol}}(n_{\text{ol}}-1)t_0 \sim 1 $ by interactions, giving rise to the characteristic time scale $ t_0 $ to observe DTC's before decays. Here $ n_{\text{ol}} $ is the density overlap for two scar DTCs hypothetically left alone in a lattice. Then, one can predict that larger distance gives a smaller density overlap  $ n_{\text{ol}} $, which prolongs the scar DTC lifetime. Such expectations are verified numerically in SM~\cite{supp} for two lattice settings relevant to the Berkeley platform. It confirms the possibility of observing DTC signatures with finite filling fractions over the experimentally accessible time, and further point out theoretically the controlling parameter for DTC lifetime therein: the distance of initially populated cells.


	{\em\color{blue} \noindent Conclusion ---} We show a distinct DTC phenomenon enforced by the analytically discovered FBS's. Its intrinsic Floquet and many-body nature stabilizes spectral pairings against translation-invariant bilinear perturbations.
	Moreover, the new scheme of checking Floquet emergent degeneracy and scar spectral pairing indicates a possible procedure to unveil the long-sought universal mechanism behind clean DTCs in arbitrary dimensions. It is also tantalizing to incorporate more intricate crystalline spacegroup symmetries aside translations into designing DTCs with unique structures and phenomena in clean systems.

	{\noindent\em\color{blue} Acknowledgment} --- This work is supported by  the National Natural Science Foundation of China Grant No. 12174389 (BH), the NSF Grant No. PHY-1806362 (THL, DS), the MURI-ARO Grant No. W911NF17-1-0323 through UC Santa Barbara (BH, THL, DS, WVL),  AFOSR Grant No. FA9550-16-1-0006 (WVL), and the Shanghai Municipal Science and Technology Major Project (Grant No. 2019SHZDZX01) (WVL).

	%

	
	\pagebreak
	
	\widetext
	\clearpage
	
	\setcounter{equation}{0}
	
	\begin{center}
		{\bf\Large Supplemental materials: Discrete time crystals enforced by Floquet-Bloch scars}
	\end{center}
	
	\renewcommand{\thesection}{S-\arabic{section}}
	\renewcommand{\theequation}{S\arabic{equation}}
	\setcounter{equation}{0}  
	\renewcommand{\thefigure}{S\arabic{figure}}
	\setcounter{figure}{0}  

	\tableofcontents

	\section{Floquet perturbation treatment}
	In this section, we will derive a strongly-driven Floquet perturbation theory for analyzing the scar stability. A concrete series up to the second order is firstly presented for intuition. Then, we would obtain the formal structure to all higher-orders and show the spectral pairing rigidity for FBS.

	\subsection{Preliminary: Factoring out perturbations}\label{sec:factor}
	The Floquet driving in Eq.~(1) of the main text has the form
	\begin{align}
		U_F =  \hat{U}_2 \hat{U}_1 =  e^{-i\hat{H}_2T/2\hbar} e^{-i (\hat{h}_1 + \lambda \hat{h'})},
	\end{align}
	where $ \hat{h}_1+\lambda \hat{h'} = \hat{H}_1 T/2\hbar  $. For later analysis, it will be convenient to factor out the perturbation related to $ \lambda \hat{h'} $ through the Baker-Campbell-Hausdorff (BCH) formula,
	\begin{align}
		\hat{U}_F &=
		\left( e^{-i\hat{H}_2T/2\hbar} e^{-i\hat{h}_1} \right)  \left( e^{i\hat{h}_1} e^{-i(\hat{h}_1+\lambda \hat{h}')} \right) \equiv \hat{U}_0 \hat{U}',
		\\ \nonumber
		\hat{U}' &= e^{i\hat{h}_1} e^{-i(\hat{h}_1+\lambda \hat{h}')} = e^{(i\hat{h}_1 - i(\hat{h}_1+\lambda \hat{h}') )
			+ \frac{1}{2} [i\hat{h}_1, -i(\hat{h}_1 + \lambda \hat{h}')] 
			+ \frac{1}{12} [i\hat{h}_1, [i\hat{h}_1, -i(\hat{h}_1 + \lambda \hat{h}')]]
			+ \frac{1}{12} [-i(\hat{h}_1 + \lambda \hat{h}'), [-i(\hat{h}_1 + \lambda \hat{h}'), i\hat{h}_1] ] + \dots
		}
		\\ \label{eq:temp1}
		&=
		e^{i\lambda (-\hat{h}' + \frac{-i}{2} [\hat{h}_1, \hat{h}'] 
			+ \frac{1}{6} [\hat{h}_1, [\hat{h}_1, \hat{h}']] + \dots) +  O(\lambda^2)}.
	\end{align}
	In our case, both $ \hat{h}_1 $ and $ \hat{h}' $ consist of bilinear terms, so their mutual commutators to any orders are also of bilinear forms. 
	Meanwhile, on the exponential part of Eq.~\eqref{eq:temp1}, all remaining terms involve commutators between $ i\hat{h}_1 $ and $ -i(\hat{h}_1+\lambda \hat{h}'d) $, and hence are at least of the order $ \lambda^1 $. Therefore, we can group all terms into an effective perturbing static Hamiltonian
	\begin{align}\label{eq:temp2}
		& \hat{U}' = e^{i\hat{h}_1} e^{-i(\hat{h}_1+\lambda \hat{h}')} =  e^{i\lambda \hat{H}'}, \qquad
		\hat{H}' = \sum_{\boldsymbol{r}\mu, \boldsymbol{r}'\mu'} J_{\boldsymbol{r}\mu, \boldsymbol{r}'\mu'} \hat{\psi}^\dagger_{\boldsymbol{r}\mu} \hat{\psi}_{\boldsymbol{r}'\mu'} = \hat{H}'_d + \hat{H}'_{nd}\\
		&
		\hat{H}_d = \sum_{\boldsymbol{r}\mu} J_{\boldsymbol{r}\mu, \boldsymbol{r}\mu} \hat{\psi}_{\boldsymbol{r}\mu}^\dagger \hat{\psi}_{\boldsymbol{r}\mu} = \sum_{\boldsymbol{r}\mu} J_{\boldsymbol{r}\mu, \boldsymbol{r}\mu} \hat{n}_{\boldsymbol{r}\mu},
		\qquad
		\hat{H}_{nd} = \sum_{\boldsymbol{r}\mu} J_{\boldsymbol{r}\mu\ne \boldsymbol{r}'\mu'} \hat{\psi}_{\boldsymbol{r}\mu}^\dagger \hat{\psi}_{\boldsymbol{r}\mu}
	\end{align}
	For our purposes, we would further factor out the onsite energy offsets in $ \hat{H}'_d $, as these terms can be grouped into $ \hat{U}_0 $:
	\begin{align}
		\hat{U}_0 e^{i\lambda \hat{H}'_d} = \hat{U}_2e^{-i\hat{h}_1} e^{i\lambda \sum_{\boldsymbol{r}\mu} J_{\boldsymbol{r}\mu, \boldsymbol{r}\mu} \hat{\psi}_{\boldsymbol{r}\mu}^\dagger \hat{\psi}_{\boldsymbol{r}\mu} } e^{i\hat{h}_1} e^{-i\hat{h}_1} 
		=
		\hat{U}_2 e^{i\lambda \sum_{\boldsymbol{r}\mu} J_{\boldsymbol{r}\mu, \boldsymbol{r}\mu} \hat{\psi}_{\boldsymbol{r},\mu+1}^\dagger  \hat{\psi}_{\boldsymbol{r}\mu+1} } e^{-i\hat{h}_1} = \hat{\tilde{U}}_0.
	\end{align}
	Here, we use the fact that under the unperturbed $ \hat{h}_1 $, $ e^{-i\hat{h}_1} \hat{\psi}_{\boldsymbol{r}\mu} e^{i\hat{h}_1} = \hat{\psi}_{\boldsymbol{r},\mu+1} $, and that onsite terms $ \sim n_{\boldsymbol{r}\mu} $ commute with the Hubbard interaction and chemical potentials in $ \hat{U}_2 $. Then, $ \hat{\tilde{U}}_0 $ is related to $ \hat{U}_0 $ by a renormalized chemical potential $ \theta_\mu \rightarrow \theta_\mu - \lambda J_{\boldsymbol{r}\mu-1, \boldsymbol{r}\mu-1} $. To factor out onsite terms, we can perform an iteration: 
	\begin{enumerate}
		\item 
		The perturbing Hamiltonian $ \hat{H}' $ can be obtained from Eq.~\eqref{eq:temp2}, and then one could obtain the diagonal terms $ \hat{H}_d' $.
		\item 
		Factor out onsite terms up to $ \lambda^1 $ by $ \hat{U}' = e^{i\lambda \hat{H}_d'} \left( e^{-i\lambda \hat{H}_d'}e^{i\lambda (\hat{H}_d' + \hat{H}_{nd})} \right) = e^{i\lambda \hat{H}_d'} e^{i\lambda \hat{H}_{nd}' + i\lambda^2 (\hat{H}_{d}^{(2)} + \hat{H}_{nd}^{(2)})} $. Here we have used the BCH formula in the last step, where $ \lambda \hat{H}_d' $ in the exponential part are canceled, and the remaining terms other than $ \lambda \hat{H}'_{nd} $ would involve commutators between $ \lambda \hat{H}_{d}' $ and $ \lambda\hat{H}_{nd}' $ and therefore are at least of the order $ \lambda^2 $.
		\item 
		Iterating the process in step 2 for $ n $ times will result in $ \hat{U}' =  e^{i \left(\lambda \hat{H}_d + \sum_{m=2}^{n-1} \lambda^m \hat{H}_d^{(m)} \right)}  e^{i \left(\lambda \hat{H}_{nd}' + \sum_{m=2}^{n-1} \lambda^m \hat{H}_{nd}^{(m)} + \lambda^n(\hat{H}_d^{(n)} + \hat{H}_{nd}^{(n)})  \right) } $. That means one can factor out the diagonal energy offset terms up to arbitrary accuracy.
	\end{enumerate}
	
	In summary, for a bilinear type of Hamiltonian, one can transform the perturbation in Eq.~(1) of the main text concerning $ \hat{H}_1 $ such that the Floquet unitary reads $ \hat{U}_F = \hat{U}_0 \hat{U}' $. Here, $ \hat{U}_0 $ is the unperturbed Eq.~(1) with $ \lambda = 0 $ and onsite chemical potentials renormalized. $ \hat{U}' = e^{i\lambda \hat{H}'} $ contains all the perturbations with $ \hat{H}' = \sum_{\boldsymbol{r}\mu \ne \boldsymbol{r}\mu} J_{\boldsymbol{r}\mu, \boldsymbol{r}\mu} \hat{\psi}_{\boldsymbol{r}\mu}^\dagger \hat{\psi}_{\boldsymbol{r}'\mu'} $.

	\subsection{Formalism and explicit results up to the second order}
	Consider a Floquet operator involving a strongly-driven but exactly solvable part $ U_0 $, and a perturbation $ U_1 $,
	\begin{align}\label{eq:expansionu1}
		U_F = U_0U',
		\qquad
		U_0|\omega_n\rangle = e^{i\omega_n} |\omega_n\rangle, \quad U' = e^{i\lambda H'} = \sum_{\alpha=0}^\infty \frac{(i\lambda)^\alpha}{\alpha!} (H')^\alpha
	\end{align}	
	Note that $ U_0 $ could be far from an identity operator unlike conventional high-frequency expansions around static limits. Now, to solve the eigenproblem perturbatively, 
	\begin{align}
		U_F |\tilde{\omega}_n\rangle = e^{i\tilde{\omega}_n} |\tilde{\omega}_n \rangle,
	\end{align}
	we expand the quasienergy and Floquet eigenstates into series
	\begin{align}\label{eq:es}
		e^{i\tilde\omega_n} 
		= e^{i\omega_n} \exp\left( i
		\sum_{\alpha=1}^\infty \lambda^\alpha \omega_n^{(\alpha)}
		\right) ,
		\qquad\qquad
		|\tilde{\omega}_n\rangle &= \dots e^{i\lambda^3 S_3} e^{i\lambda^2 S_2} e^{i\lambda S_1}|\omega_n\rangle,
	\end{align}
	such that $ e^{i\lambda^\alpha S_\alpha} $ should diagonalize $ U_F $ up to the order of $ \lambda^\alpha $. 
	Note that unlike previous perturbation treatments~\cite{Else2016} expressing $ (\alpha+1) $-th order results by $ \alpha $-th order ones, we would aim at obtaining series of any orders in terms of the zeroth-order solutions $ \omega_n, |\omega_n\rangle $. This is necessary for quantitative evaluations like spectral pairing rigidity, as only the zeroth-order results are exactly solvable. 
	
	To $ \lambda^1 $ order,
	\begin{align}\nonumber
		\langle \omega_m | e^{-i\lambda S_1} U_0 e^{i\lambda H'}  e^{i\lambda S_1} |\omega_n \rangle
		= &\,
		Q^{(1)}_n \left(
		\langle \omega_m |U_0|\omega_n\rangle  + i\lambda \langle \omega_m | ( U_0 H' + [U_0,S_1]) |\omega_n\rangle  
		\right)
		+ O(\lambda^2) \\ \nonumber
		= &\,
		Q^{(1)}_n \left(
		e^{i\omega_m}\delta_{mn} + i\lambda (  e^{i\omega_m} [H']_{mn} + (e^{i\omega_m} - e^{i\omega_n} ) [S_1]_{mn}) 
		\right)
		+ O(\lambda^2).\\ \label{eq:pert1}
		= &\,
		Q^{(1)}_n e^{i\omega_m}\left(
		\delta_{mn} + i\lambda ( [H']_{mn} + (1 - e^{i\omega_{nm}} ) [S_1]_{mn}) 
		\right)
		+ O(\lambda^2).
	\end{align}
	where $ \omega_{nm} = \omega_n - \omega_m $ and matrix elements i.e. $ [S_1]_{mn} = \langle \omega_m |S_1|\omega_n\rangle $. The normalization constant $ Q^{(1)}_n $ makes sure that the diagonal term ($ m=n $) on the right-hand-side for perturbed eigenvalue, after keeping terms only up to $ \lambda^1 $, is still unitary. 
	To diagonalize $ U_F $ up to $ \lambda^1 $, choose
	\begin{align}\label{eq:s1}
		[S_1]_{mn} = 
		\frac{ [H']_{mn} }{e^{i\omega_{nm}}-1}, 
		\quad
		(m\ne n); 
		\qquad\qquad
		[S_1]_{nn}= 0.
	\end{align}
	Then, compare the diagonal term in Eq.~\eqref{eq:pert1} with Eq.~\eqref{eq:es}, we have a similar result as in static case
	\begin{align}
		\omega_n^{(1)} = [H']_{nn}.
	\end{align}
	
	A qualitative difference from static perturbation can be observed in the $ \lambda^2 $ order,
	\begin{align}\nonumber
		& \langle \omega_m |e^{-i\lambda S_1} e^{-i\lambda^2S_2}  U_0 e^{i\lambda H'}  e^{i\lambda^2 S_2} e^{i\lambda S_1} |\omega_n\rangle
		\\ \label{eq:detail2}
		=&\,
		Q_n^{(2)} \left(
		e^{i\omega_n}(1+i\lambda[H']_{nn})\delta_{mn} +
		\lambda^2 \langle \omega_m | 
		[
		-\frac{1}{2}(U_0S_1^2 + S_1^2U_0) + i[U_0, S_2] - \frac{1}{2} U_0 (H')^2
		+ [S_1, U_0H'] + S_1U_0S_1
		]
		|\omega_n\rangle
		\right) + O(\lambda^3).
	\end{align}
	To diagonalize it, we similarly require the $ \lambda^2 $ term for off-diagonal elements $ m\ne n $ to vanish,
	\begin{align}
		\nonumber
		0= & \,
		\lambda^2 \left(
		-\frac{[S_1^2]_{mn}}{2} (e^{i\omega_m} + e^{i\omega_n}) + i[S_2]_{mn} (e^{i\omega_m} - e^{i\omega_n} )  
		- \frac{[(H')^2]_{mn}}{2} e^{i\omega_m} - [H'S_1]_{mn} e^{i\omega_m} + \sum_l [S_1]_{ml}[H']_{ln} e^{i\omega_l}
		+ \sum_l [S_1]_{ml}[S_1]_{ln} e^{i\omega_l} 
		\right) ,
	\end{align}
	A straightforward calculation simplifies it to
	\begin{align}\nonumber
		i[S_2]_{mn}(e^{i\omega_{nm}}-1) 
		&= \sum_l \frac{[H']_{ml} [H']_{ln}}{(e^{i\omega_{lm}}-1) (e^{i\omega_{nl}}-1)} \left(
		-\frac{1+e^{i\omega_{nm}}}{2} 
		-\frac{(e^{i\omega_{lm}}-1) (e^{i\omega_{nl}}-1)}{2}
		-(e^{i\omega_{lm}}-1) 
		+e^{i\omega_{lm}} (e^{i\omega_{nl}}-1) + e^{i\omega_{lm}}
		\right)
		\\ \label{eq:detail3}
		&= \sum_l \frac{[H']_{ml} [H']_{ln}}{(e^{i\omega_{lm}}-1) (e^{i\omega_{nl}}-1)} \frac{e^{i\omega_{nl}} - e^{i\omega_{lm}}}{2}
	\end{align}
	Therefore, we can similarly choose
	\begin{align}
		[S_2]_{mn} &= \frac{-i}{e^{i\omega_{nm}}-1}  \sum_l \frac{[H']_{ml} [H']_{ln}}{ (e^{i\omega_{lm}}-1) (e^{i\omega_{nl}}-1) }  
		\frac{e^{i\omega_{nl}} -e^{i\omega_{lm}} }{2} , \quad
		(m\ne n);
		\qquad\qquad
		[S_2]_{mm} = 0,
	\end{align}
	Then, the diagonal term of Eq.~\eqref{eq:detail2}, where $ \lambda^2 $ terms corresponding to the right-hand-side of Eq.~\eqref{eq:detail3} with $ m=n $, gives the second order correction to quasienergy
	\begin{align}\label{eq:2ndorder}
		{\color{blue}
			\omega_n^{(2)} } &= - \sum_{l\ne n} \left| \frac{[H']_{nl}}{ (e^{i\omega_{ln}}-1) }  \right|^2 \sin\omega_{ln} 
		=
		{\color{blue}
			- \frac{1}{2} \sum_{l\ne n} | [H']_{nl} |^2 \cot \frac{\omega_{ln}}{2}
		}
	\end{align}
	We can benchmark the results by considering high frequency limits where all $ \omega_{nl} $'s are very small. Then, we can expand $ \cot(\omega_{nl}/2)\approx 2/\omega_{nl} $, so  $\omega_n^{(2)} = \sum_{l\ne n} |[H']_{nl}|^2/\omega_{nl}$, which recovers the familiar static perturbation result. 
	
	It is worth noting that in the Floquet case, level differences $ \omega_{ln} $ contribute a periodic $ 2 \pi $ correction $ \cot(\omega_{ln}/2) $ in contrast to the $ 1/\omega_{ln} $ factor for static cases without periodicity. Such a difference is crucial for the spectral pairing rigidity only possible in Floquet systems.
	
	In sum, up to the second order, the perturbed quasienergy reads
	\begin{align}\label{eq:pert2}
		{\color{blue}
			e^{\tilde{\omega}_n} \approx e^{i\omega_n} \exp\left( i\lambda[H']_{nn} - i \frac{\lambda^2}{2} \sum_{l\ne n} |[H']_{nl}|^2 \cot\frac{\omega_{ln}}{2} \right),
		}
	\end{align}
	For completeness, we also write the dressed eigenstates
	\begin{align}\nonumber
		|\tilde{\omega}_n\rangle & \approx (1+i\lambda^2 S_2) (1+i\lambda S_1 -\frac{\lambda^2}{2}S_1^2) |\omega_n\rangle\\
		\nonumber
		&=
		|\omega_n\rangle + i\lambda \sum_{m\ne n} [S_1]_{mn} |\omega_m\rangle  
		+ \lambda^2 \left( 
		\sum_m \sum_{l\ne m, n} -\frac{[S_1]_{ml} [S_1]_{ln}}{2}
		+ i[S_2]_{mn}
		\right) |\omega_m\rangle.
		\\ \label{eq:perturbedfbseigs}
		&= |\omega_n\rangle  
		+ i\lambda \sum_{m\ne n} \frac{[H']_{mn}}{e^{i\omega_{mn}} -1} |\omega_m\rangle 
		- \frac{\lambda^2}{2}  \left(
		\sum_m \sum_{l\ne m,n} [S_1]_{ml} [S_1]_{ln}  
		+ \sum_{m\ne n} \sum_{l\ne m,n}  \frac{e^{i\omega_{ml}} - e^{i\omega_{ln}}}{e^{i\omega_{mn}} -1}  [S_1]_{ml} [S_1]_{ln}
		\right) |\omega_m\rangle 
	\end{align}
	up to a normalization factor. Here $ [S_1]_{mn} $ etc are given by Eq.~\eqref{eq:s1}

	\subsection{Scar spectral pairing rigidity up to the second order}
	Now, we apply the previously developed strong-drive Floquet perturbation theory to analyze the scar spectral pairings. Recall Eq.~(3) (4) in the main text for the eigenstates at $ \lambda=0 $, which we will take as eigenstates for $ U_0 $,  
	\begin{align}\tag{2}
		\left| \boldsymbol{k},  \ell, \{n_{\boldsymbol{r}\mu}\} \right\rangle 
		&=\,
		\frac{1}{\sqrt3} \sum_{m=0,1,2} e^{-i(\frac{2\pi m}{3}\ell -\alpha_m)} 
		|\boldsymbol{k}, \{n_{\boldsymbol{r},\mu+m \text{ mod } 3} \} \rangle,\\
		\tag{3}
		E\left(\ell, \{n_{\boldsymbol{r}\mu}\}\right) &=  \frac{2\pi}{3} \ell  + \phi_2 \sum_{\boldsymbol{r}\mu} n_{\boldsymbol{r} \mu} (n_{\boldsymbol{r}\mu} - 1),  \qquad
		\ell = 0, \pm 1.
	\end{align}
	Each level labeled by $ |\boldsymbol{k},\ell,\{n_{\boldsymbol{r}\mu}\}\rangle $ corresponds to the zeroth order $ |\omega_n\rangle $ here, and quasienergy $ e^{iE(\ell,\{n_{\boldsymbol{r}\mu}\})} $ is denoted $ e^{i\omega_n} $. Among them, the special set of scars are defined in Eq.~(4) of the main text, 
	\begin{align}\tag{4}
		|\boldsymbol{k}, \ell, N_b \rangle = \frac{1}{\sqrt{3}} \sum_{m=0,1,2} 
		e^{-i(\frac{2\pi m}{3}\ell - \alpha_m)} 
		|\boldsymbol{k},\{n_{\boldsymbol{r}\mu} =  \delta_{\boldsymbol{r},\boldsymbol{0}} \delta_{\mu,m}N_b \} \rangle,\qquad
		E_{\text{scar}}(\ell) = \frac{2\pi\ell}{3}+\phi_2 N_b(N_b-1),\qquad
		\ell = 0, \pm1.
	\end{align}
	In this subsection, we would aim at proving that all the three scars $ \ell=0,\pm $ in each $ \boldsymbol{k} $ sector will receive the same quasienergy corrections in Eq.~\eqref{eq:pert2}, with the perturbing Hamiltonian $ H' = \sum_{\boldsymbol{r} \mu \ne \boldsymbol{r}'\mu'} J_{\boldsymbol{r}\mu, \boldsymbol{r}'\mu'} \hat{\psi}^\dagger_{\boldsymbol{r}\mu} \hat{\psi}_{\boldsymbol{r}' \mu'} $ being a generic bilinear hopping one respecting translation invariance. 
	
\begin{figure}
		[h]
		\includegraphics[width=7cm]{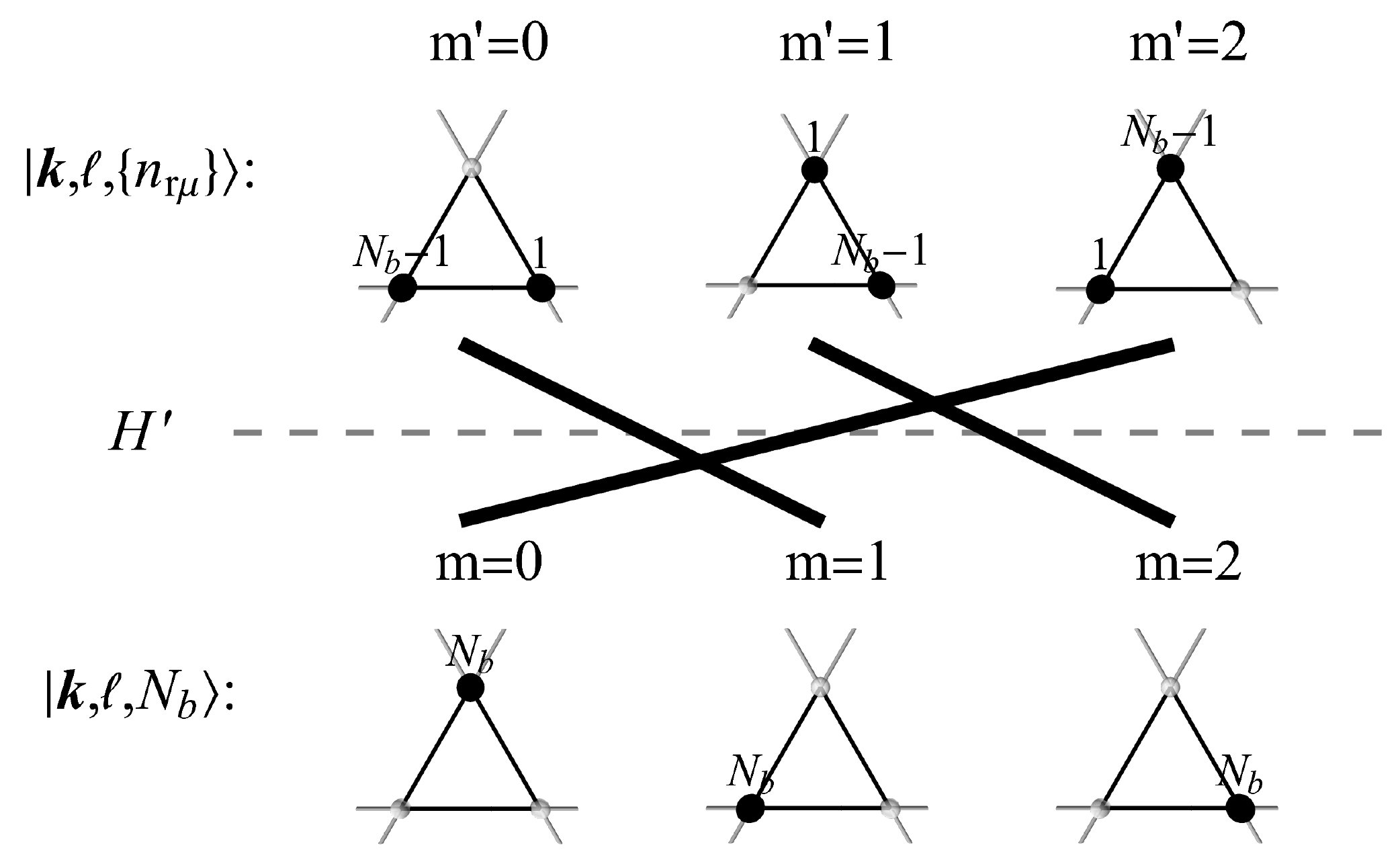}\\
		\caption{\label{fig:n5n41}Exemplary non-vanishing matrix elements in the second order perturbation for scar states $ |\boldsymbol{k},\ell,N_b\rangle $. The non-vanishing matrix elements would have $ m' = m + \Delta m $ mod $ 3 $, where $ \Delta m = 2 $ for all $ m, m' $.}
	\end{figure}

	First, as onsite chemical potential terms are all factored out into $ U_0 $ (see subsection~\ref{sec:factor}), the first order corrections $ E_\ell^{(1)} \propto \langle \boldsymbol{k},\ell, N_b| H' | \boldsymbol{k}, \ell, N_b\rangle = 0 $ vanish --- in $ |\boldsymbol{k}, \ell, N_b\rangle $ (for $ N_b>1 $) all particles are allocated onto a single site and no hopping happens. Next, for the second order corrections, we write explicitly
	\begin{align}
		E_\ell^{(2)} = -\frac{1}{2} \sum_{\{n'_{\boldsymbol{r}\mu}\}, \ell'}
		|\langle \boldsymbol{k},\ell, N_b | H' | \boldsymbol{k}, \ell', \{n'_{\boldsymbol{r}\mu}\} \rangle |^2 \cot  \left( \frac{E(\ell', \{n'_{\boldsymbol{r}\mu}\}) - E_{\text{scar}}(\ell)}{2} \right).
	\end{align}
	As the perturbing Hamiltonian only involves hopping terms, the non-vanishing matrix elements have $ \{n_{\boldsymbol{r}\mu}'\} $ with $ (N_b-1) $ particles on one site, and $ 1 $ particle on another. Written in the main text notation, $ \{n_{\boldsymbol{r}\mu}'\} \in Q_a = \{(q^{(a)}_1,N^{(a)}_1)=(1,N_b-1), (q^{(a)}_2,N^{(a)}_2)=(1,1), (q^{(a)}_3,N^{(a)}_3)=(3L^2-2,0) \} $. Thus, $ E(\ell', \{n'_{\boldsymbol{r}\mu}\}) = 2\pi\ell'/3 + \phi_2\sum_j q^{(a)}_jN^{(a)}_j(N^{(a)}_j-1) =  2\pi\ell'/3 + \phi_2(N_b-1)(N_b-2) $, and all
	\begin{align}
		E(\ell',\{n_{\boldsymbol{r}\mu}\}) - E_{\text{scar}}(\ell) = - (2\pi(\ell-\ell')/3 + 2\phi_2).
	\end{align}
	Further, within the manifold $ \{n_{\boldsymbol{r}\mu}'\} \in Q_a $,
	\begin{align}
		\langle \boldsymbol{k},\ell, N_b | H' | \boldsymbol{k}, \ell', \{n'_{\boldsymbol{r}\mu}\} \rangle 
		&=
		\frac{1}{3}\sum_{m,m'=0}^2 \langle \boldsymbol{k}, \{n_{\boldsymbol{r}\mu} = \delta_{\boldsymbol{r},\boldsymbol{0}} \delta_{\mu,m} N_b\}| H' | \boldsymbol{k}, \{n'_{\boldsymbol{r}\mu+m'}  \}\rangle e^{i\frac{2\pi}{3}(m\ell - m'\ell')} e^{-i(\alpha_m - \tilde\alpha_{m'})},
	\end{align}
	the site with $ N_b $ particles for $ \{n_{\boldsymbol{r}\mu}\} $ must be the same as the site with $ (N_b-1) $ particles for $ \{n_{\boldsymbol{r}\mu}'\} $, see Fig.~\ref{fig:n5n41} for example. 
	That means a non-vanishing term would have $ \langle \boldsymbol{k}, \{n_{\boldsymbol{r}\mu} = \delta_{\boldsymbol{r},\boldsymbol{0}} \delta_{\mu,m} N_b\}| H' | \boldsymbol{k}, \{n'_{\boldsymbol{r}\mu+m'}  \}\rangle  \propto \delta_{m', m+\Delta m}  $, with a fixed $ \Delta m $ (mod 3) for all $ m, m' $. Then,
	\begin{align}\nonumber
		|\langle \boldsymbol{k},\ell, N_b | H' | \boldsymbol{k}, \ell', \{n'_{\boldsymbol{r}\mu}\} \rangle |^2
		&=
		\left|\frac{1}{3}  e^{-i\frac{2\pi \Delta m}{3}\ell'}
		\sum_{m=0}^2 \langle \boldsymbol{k}, \{n_{\boldsymbol{r}\mu} = \delta_{\boldsymbol{r},\boldsymbol{0}} \delta_{\mu,m} N_b\}| H' | \boldsymbol{k}, \{n'_{\boldsymbol{r}\mu+m+\Delta m}  \}\rangle e^{i\frac{2\pi}{3}m(\ell - \ell')} e^{-i(\alpha_m - \tilde\alpha_{m'})}
		\right|^2.
	\end{align}
	In sum, the second order quasi-energy correction
	\begin{align}
		E_\ell^{(2)} &= \frac{1}{18} \sum_{ \{n_{\boldsymbol{r}\mu}'\},\ell'} 
		\left|
		\sum_{m=0}^2 \langle \boldsymbol{k}, \{n_{\boldsymbol{r}\mu} = \delta_{\boldsymbol{r},\boldsymbol{0}} \delta_{\mu,m} N_b\}| H' | \boldsymbol{k}, \{n'_{\boldsymbol{r}\mu+m+\Delta m}  \}\rangle e^{i\frac{2\pi}{3}m{\color{red}(\ell - \ell')}} e^{-i(\alpha_m - \tilde\alpha_{m'})}
		\right|^2
		\cot\left(
		\frac{\pi{\color{red}(\ell-\ell')}}{3} + \phi_2
		\right)
	\end{align}
	indeed only depend on the difference $ (\ell - \ell') $, but not individual $ \ell $ and $ \ell' $. Thus, we can generically write $ E_{\ell_1}^{(2)} \equiv \sum_{\ell'} \varepsilon(\ell_1-\ell') = \sum_{\ell'} \varepsilon(\ell_2 - (\ell'-\ell_1+\ell_2)) = \sum_{\ell''} \varepsilon(\ell_2-\ell'') = E_{\ell_2}^{(2)} $, proving the spectral pairing rigidity up to the second order.

	\subsection{Numerical verification}
	
\begin{figure}
		[th]
		\parbox[b]{6cm}{
			\includegraphics[width=5.5cm]{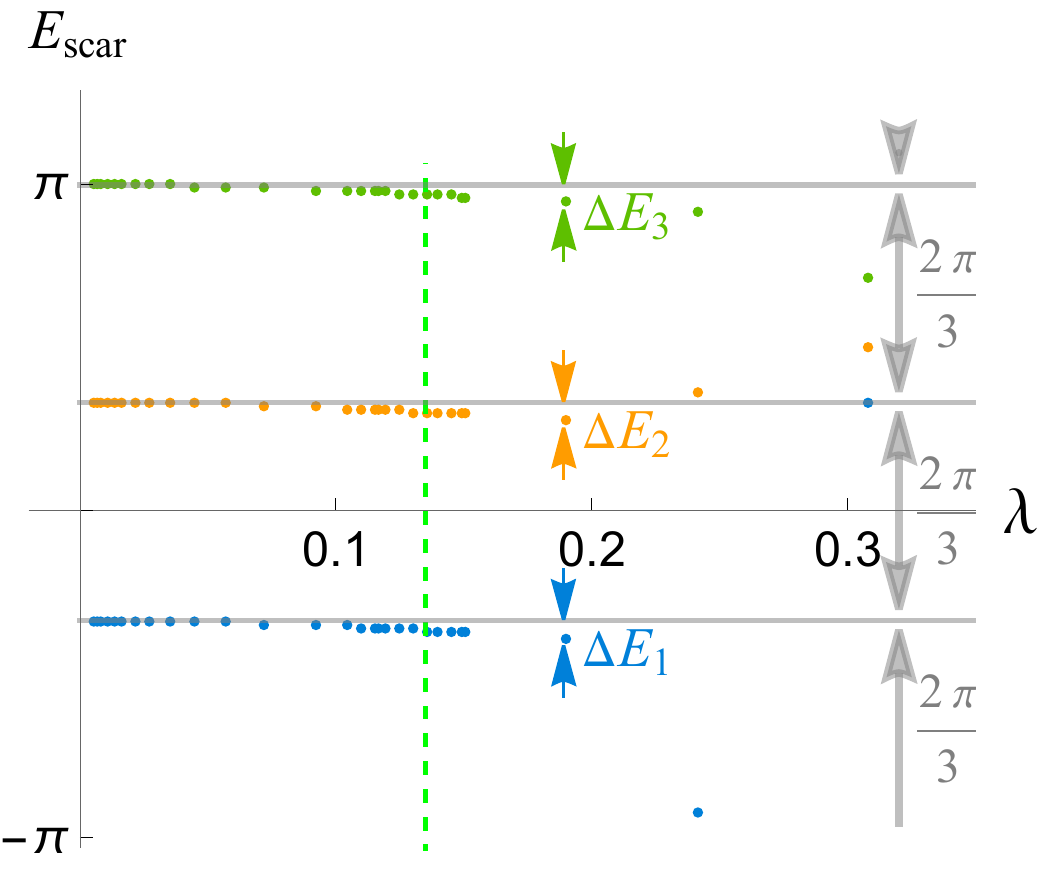}\\
			(a) Quasienergy for largest IPR eigenstates
		}
		\parbox[b]{6cm}{
			\includegraphics[width=5.5cm]{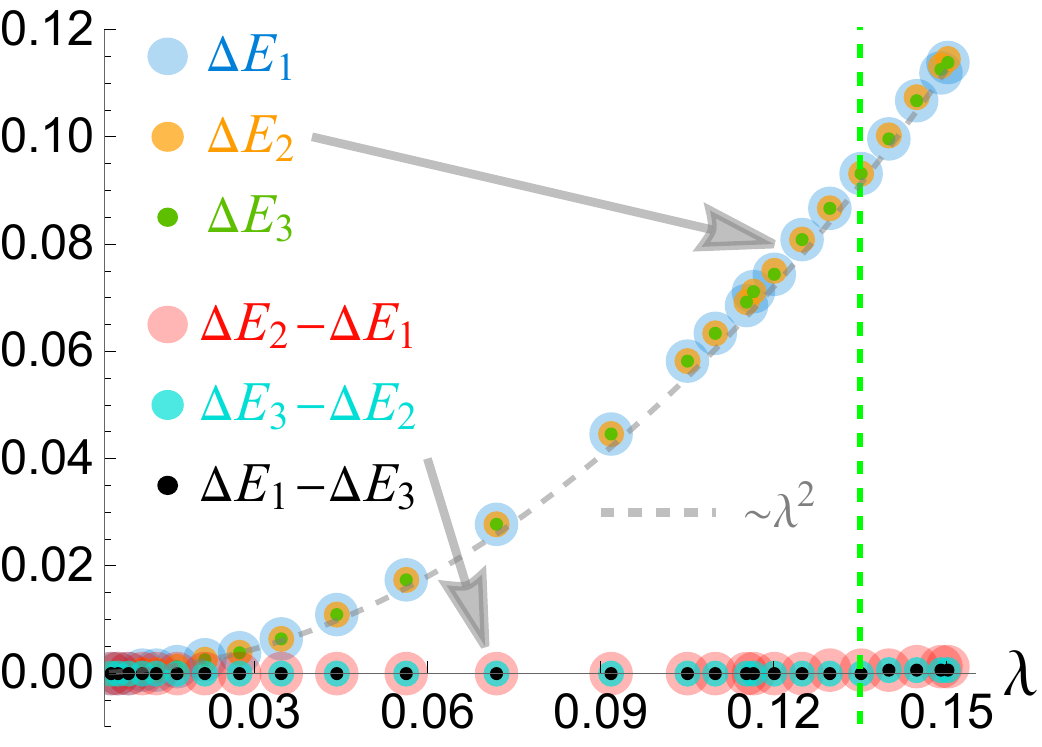}\\
			(b) Deviations and spectral pairing rigidity
		}
		\parbox[b]{5.5cm}{
			\includegraphics[width=4.5cm]{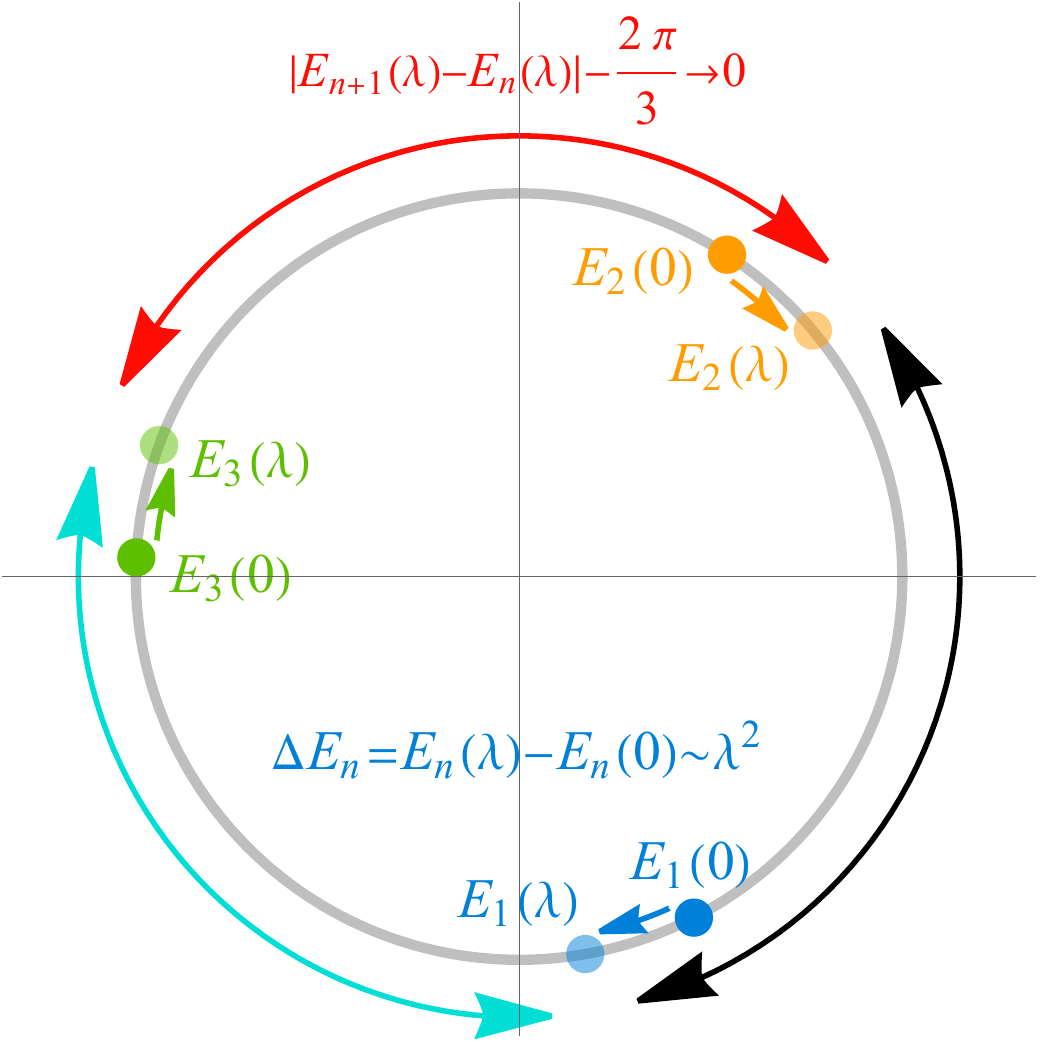}\\
			(c) Spectral pairing for FBS's
		}
		\caption{\label{fig:smpairing} Test of spectral pairing for the main text model. (a) Based on the data of Fig.~(3) in the main text, we obtain the quasi-energy of the 3 eigenstates with largest IPR's. Below the transition $ \lambda\approx0.135 $ indicated by the green dashed line, they correspond to the FBS's, while for $ \lambda >0.135 $ all eigenstates have similarly vanishing IPR's. We take $ \lambda=0.005 $ as a reference point for $ \lambda\rightarrow0 $ limits (corresponding to gray horizontal lines), and measure quasienergy deviations  $ \Delta E_n \equiv  E_n(0.005) -E_n( \lambda) $. (b) The spectral deviations of each FBS. We see that {\em individual} scar quasienergy $ \Delta E_n $ indeed exhibits a $ \sim\lambda^2 $ deviation of Fermi-Golden rule type. However, different scars demonstrate almost identical  deviations $ \Delta E_{n_1} - \Delta E_{n_2} \rightarrow 0 $, such that their mutual spacing of $ 2\pi/3 $ in (a) is preserved, leading to a stable $ 2\pi T/ (2\pi/3) = 3T $ periodic DTC oscillations. (c) Schematic illustrate for scar spectral pairings. All data is for the $ \boldsymbol{k}=\boldsymbol{0} $ sector, and other $ \boldsymbol{k} $ sectors demonstrate essentially the same characters. $ L=3 $, and parameters are the same as Fig.~1 in the main text.}
	\end{figure}

	Now we verify the previous analysis numerically. First, we test the model in Eq.~(1) of the main text. Quasienergy for the 3 scar states, obtained according to maximal momentum space IPRs as in Fig.~3 of the main text, is shown in Fig.~\ref{fig:smpairing} (a). At the anchor point $ \lambda\rightarrow0 $, mutual spacing for FBS's approaches $ 2\pi/3 $ giving the $ 3T $-periodic oscillations. Then, under perturbations $ \lambda $, each individual FBS indeed receive an energy correction dominated by $ \sim\lambda^2 $ as expected. However, the key feature is that all three FBS's receive identical quasienergy corrections, as shown by Fig.~\ref{fig:smpairing} (b), leading to a rigid $ 2\pi/3 $ spectral pairing between pairs of FBS's schematically illustrated in Fig.~\ref{fig:smpairing} (c).

\begin{figure}
		[thb]
		\parbox[b]{3cm}{
			\includegraphics[width=2cm]{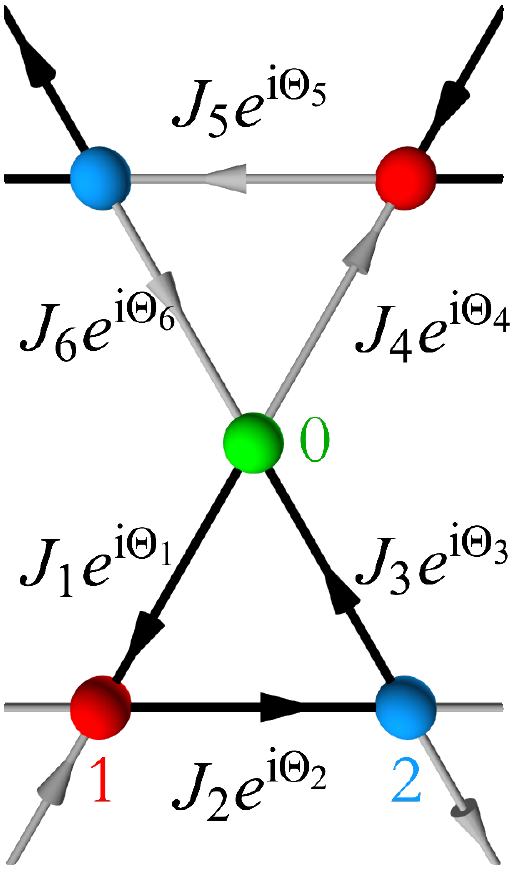}\\
			(a) Random bonds
		}
		\parbox[b]{10cm}{
			\includegraphics[width=7.5cm]{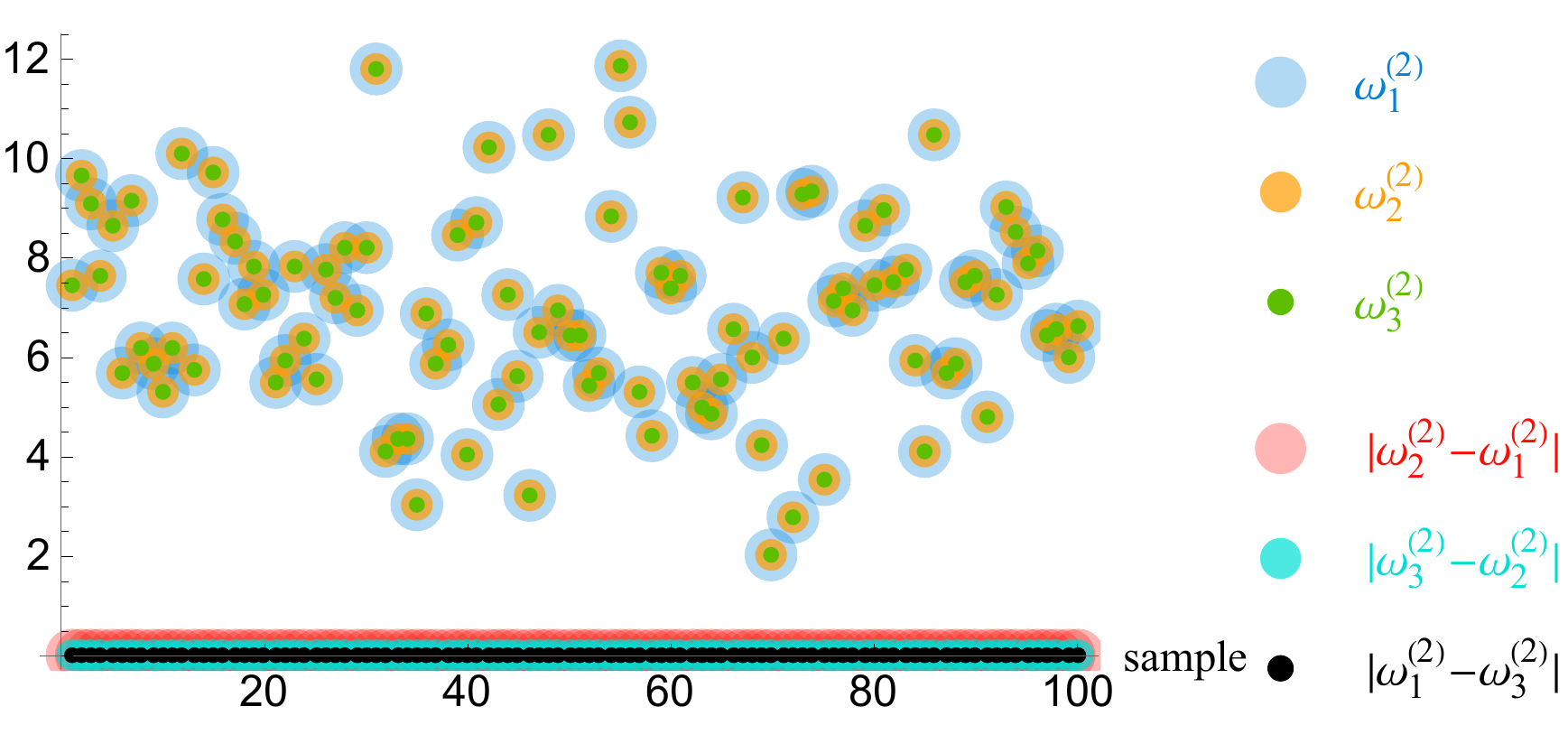}\\
			(a) Second order quasienergy corrections (in units of $ \lambda^2 $)
		}
		\caption{\label{fig:2nd}Generalized model allowing for random bond parameters (preserving translation symmetry) up to nearest neighbor hoppings. 
			Floquet operator takes the form in Eq.~\eqref{eq:expansionu1}, where $ N_b=5, L=3 $, $ U_0 $ is for the main text model at the anchor point $ \phi_1=2\pi/(3\sqrt{3}), \lambda=0 $, and the perturbation $ U'=e^{i\lambda H'} $ . (a) Random hopping parameters for $ H' $. (b) Results for the $ \boldsymbol{k}=\boldsymbol{0} $ sector. Each sample means one set of parameters in (a). To compare with Fig.~\ref{fig:smpairing}, note the rescaling for each point, $ \omega_n^{(2)} \sim \Delta E_n/\lambda^2 $, where $ \omega_n^{(2)} $ is given by Eq.~\eqref{eq:2ndorder}.
		}
	\end{figure}

	Further, we test the spectral pairing rigidity against more generic perturbations using the second order perturbation results in Eq.~\eqref{eq:2ndorder}. Here we generalize $ H' $ to involve all possible hoppings up to nearest neighbors, where each bond can possess different hopping matrix elements as shown by Fig.~\ref{fig:2nd} (a). These bonds possess random strengths $ J_n\in[0,1] $ and also carry random fluxes $ \Theta_n\in[0,2\pi] $. Each set of $ \{(J_n,\Theta_n) |n=1,\dots,6)\} $ consists of a sample. From the results in Fig.~\ref{fig:2nd} (b) for different sets of samples, it is clear that each scar quasienergy can receive notable second order corrections $ \omega_n^{(2)} $ as given by Eq.~\eqref{eq:2ndorder}. However, all 3 scars receive equal amount of corrections $ |\omega_{n+1}^{(2)} - \omega_{n}^{(2)}| \rightarrow0 $ (up to numerical errors), such that their mutual spectral pairing remain rigidly $ 2\pi/3 $ just like that for $ U_0 $, reproducing again the scheme in Fig.~\ref{fig:smpairing} (c).

	\subsection{Higher orders}
	Formally, as can be seen from Eq.~\eqref{eq:pert2} and Eqs.~\eqref{eq:expansionu1} -- \eqref{eq:es}, the perturbed quasi-energy at the $ \alpha $-th order takes the most generic form satisfying Floquet quasienergy periodicity as
	\begin{align}\label{eq:pertalpha}
		\omega_n^{(\alpha)} =  
		\sum_\beta
		\left.\sum_{l_1\dots l_{\alpha-1}}\right.'
		g_\beta( [H']_{nl_1} [H']_{l_1l_2} \dots [H']_{l_{\alpha-2} l_{\alpha-1} } [H']_{l_{\alpha-1}n})
		h_\beta(e^{i\omega_{nl_1}}, e^{i\omega_{l_1l_2}}, \dots, e^{i\omega_{l_{\alpha-1}n}}),
	\end{align}
	where $ \beta=1,2,\dots $ denotes a set of functions. The major difference between the second and higher orders is that one would encounter, i.e. $ l_1 $ and $ l_2 $ denoting unperturbed eigenstates within the same degenerate manifold. To be concrete, we give the explicit third order result 
	\begin{align}\label{eq:energy3}
		\omega_n^{(3)} 
		&=
		\frac{1}{3!} \sum_{l_1\ne l_2\ne n} [H']_{nl_1} [H']_{l_1l_2} [H']_{l_2n} 
		\left[			
		\frac{1}{2}
		\frac{\cos(\omega_{l_1n}/2)}{\sin(\omega_{l_2l_1}/2) \sin(\omega_{nl_2}/2)} 
		+
		\frac{1}{2}
		\frac{\cos(\omega_{nl_2}/2)}{\sin(\omega_{l_1n}/2) \sin(\omega_{l_2l_1}/2)} 
		-
		\frac{\cos(\omega_{l_2l_1}/2)}{\sin(\omega_{l_1n}/2) \sin(\omega_{nl_2}/2)} 
		\right] 
	\end{align}
	Here, two different levels denoted by $ l_1, l_2 $ could be different configurations in the same degenerate manifold $ \ell, Q_a = \{(q^{(a)}_1,N^{(a)}_1)=(1,N_b-1), (q^{(a)}_2,N^{(a)}_2)=(1,1), (q^{(a)}_3,N^{(a)}_3)=(3L^2-2,0)\} $, so the two different levels $ l_1, l_2 $ both have unperturbed quasienergy $ 2\pi\ell/3 + \sum_j q_j N_j(N_j-1) = 2\pi\ell/3 + (N_b-1)(N_b-2) $. As such, we need to first perform a degenerate level perturbation for non-scar eigenstates within each subspace $ \ell, Q_a $.
	\begin{figure}
		[h]
		\includegraphics[width=7.5cm]{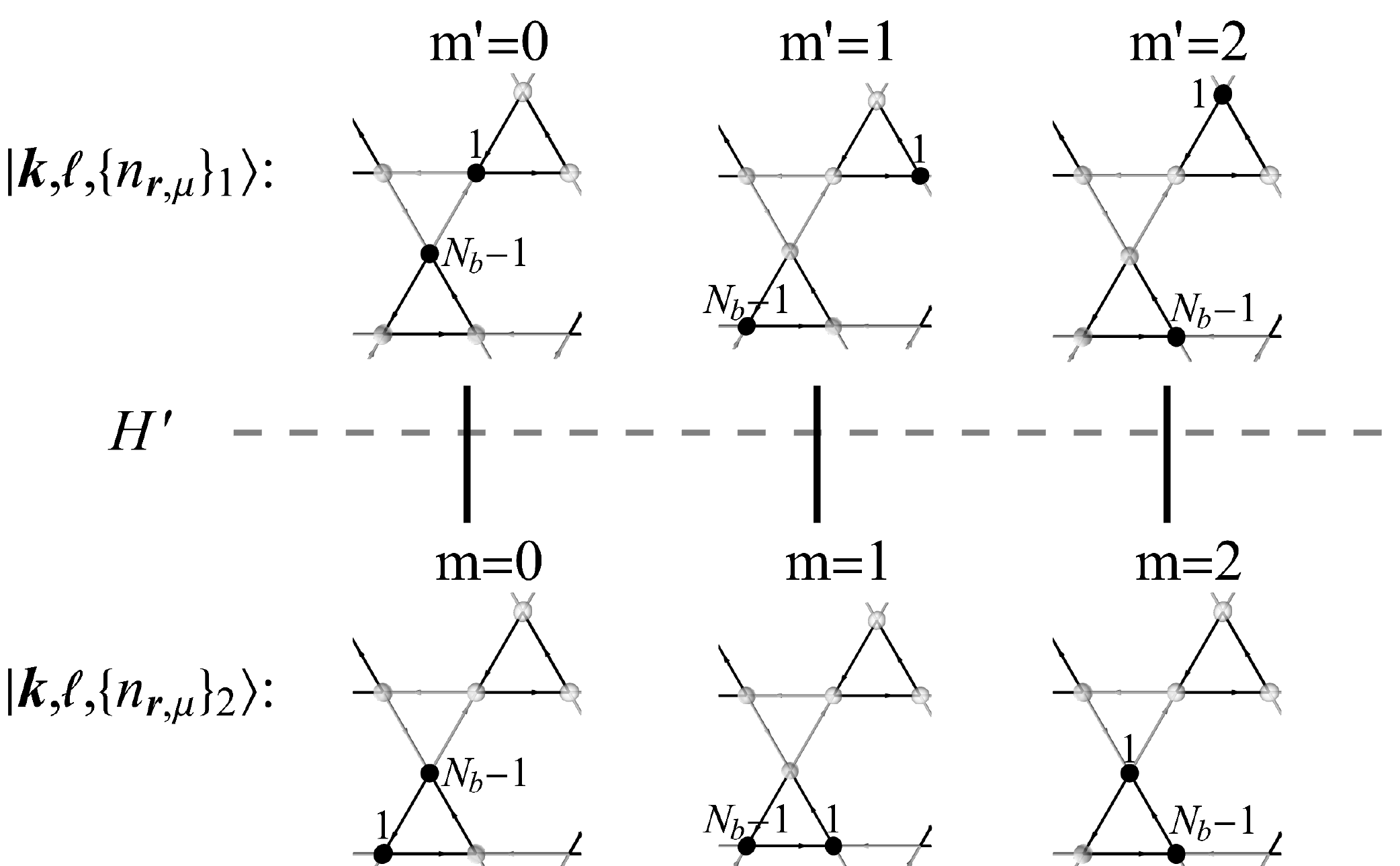}
		\caption{\label{fig:n41AB} Non-vanishing submatrix elements within the $ Q_a = \{(q^{(a)}_j, N^{(a)}_j) = (1,N_b-1), (1,1), (3L^2-2,0)\} $ degenerate manifolds. Similar to the case in Fig.~\ref{fig:n5n41}, the site with majority particle $ N_b-1 $ forces the matrix element to only depend on the relative $ \ell - \ell = 0 $ --- so there is no dependence on $ \ell $ for degenerate sub-space corrections.}
	\end{figure}
	
	Importantly, the submatrix spanned by $ \{ |\boldsymbol{k}, \ell, \{n_{\boldsymbol{r}\mu}\}\rangle | \{n_{\boldsymbol{r}\mu}\in Q_a\}\} $ would contribute a degeneracy-lifting energy {\em independent} of $ \ell $. This is due to the same reason as before: the site with $ N_b-1 $ particles must be the same in order for the hopping matrix element to be non-vanishing. One can choose the gauge that the unperturbed eigenstates within $ Q_a $ have the $ (N_b-1) $-particle site being the same for each $ m $ in different $ \{n_{\boldsymbol{r}\mu}\} $, as shown in Fig.~\ref{fig:n41AB}. Then, the degenerate space corrections in different $ \ell $ sectors are the same. That means for degenerate eigenstates within the manifold $ Q_a = \{(q_j^{(a)}, N_j^{(a)})|j=1,2,\dots,M\} $, one only needs to replace Eq.~(2), (3) in the main text with 
	\begin{align}\tag{$ 2' $}
		&
		|\boldsymbol{k},\ell, \{ \{n_{\boldsymbol{r}\mu}\} \in Q_a \}, \gamma\rangle  
		= 
		\sum_{\{n_{\boldsymbol{r}\mu}\} \in Q_a} A_{\{n_{\boldsymbol{r}\mu}\}}^{(\gamma)} |\boldsymbol{k}, \ell, \{n_{\boldsymbol{r}\mu}\} \rangle 
		=
		\frac{1}{\sqrt3} \sum_{m=0,1,2} e^{-i\frac{2\pi m}{3}\ell} 
		\left(
		\sum_{\{n_{\boldsymbol{r}\mu}\} \in Q_a} A_{\{n_{\boldsymbol{r}\mu}\}}^{(\gamma)} e^{i\alpha_m} |\boldsymbol{k}, \{n_{\boldsymbol{r}\mu}\} \rangle 
		\right)
		\\
		\tag{$ 3' $}
		&E\left(\ell,\{ \{n_{\boldsymbol{r}\mu}\} \in Q_a\},\gamma\right) = \frac{2\pi}{3} \ell + \phi_2 \sum_{j=1}^M q_j N_j(N_j-1)+ E_\gamma(\{n_{\boldsymbol{r}\mu}\}\in Q_a).
	\end{align}
	Due to the absence of $ \ell $ in the submatrix within degenerate $ Q_a $ manifold, the quasienergies in each $ \ell $ sector are lifted by the $ \ell $-independent $ E_\gamma $, and the coefficients $ A_{\{n_{\boldsymbol{r}\mu}\}}^{(\gamma)} $ are also independent of $ \ell $. 
	
	Now, as all degeneracies are lifted, and the resulting levels still has the structure of identical plethoras of $ \ell=0,\pm1 $, we can use the previous analysis to prove spectral pairing rigidity. Specifically,  in Eq.~\eqref{eq:pertalpha}, when we change the scar level  
	\begin{align}\label{eq:shift}
		n\sim (\ell, N_b) \quad \rightarrow \quad  n'\sim (\ell\pm1, N_b) 
	\end{align}
	the right-hand-side should remain the same. 
	Specifically, we note that quantum numbers $ \ell=0,\pm1 $ is only defined modulo 3. That means summing over $ \sum_{\ell=-1,0,1} $ is equivalent to $ \sum_{\ell+1=0,1,2} = \sum_{\ell+1=0,1,-1} $ and $ \sum_{\ell-1=-2,-1,0}=\sum_{1,-1,0} $. Then, Eq.~\eqref{eq:shift} would be compensated by a simultaneous change of the dummy index in Eq.~\eqref{eq:pertalpha}
	\begin{align}
		l_1\sim (\ell_1,\{n_{\boldsymbol{r}\mu}\}_1)
		\quad\rightarrow\quad
		l_1'\sim(\ell_1\pm1, \{n_{\boldsymbol{r}\mu}\}_1),
		\qquad\qquad
		l_{\alpha-1}\sim (\ell_{\alpha-1},\{n_{\boldsymbol{r}\mu}\}_{\alpha-1})
		\quad\rightarrow\quad
		l_{\alpha-1}'\sim(\ell_{\alpha-1}\pm1, \{n_{\boldsymbol{r}\mu}\}_{\alpha-1}),
	\end{align}
	while all the remaining indices are unaffected.
	
	In summary, we have proved that for perturbations $ H' $ of a generic bilinear form conserving translation symmetries, all scar levels $ |\boldsymbol{k},\ell,N_b\rangle $ will receive the same amount of energy correction in the perturbation series, leading to the spectral pairing rigidity $ |\Delta E|=2\pi/3 $ for FBS's.
	
	\section{Entanglement entropy at the anchor point $ \lambda\rightarrow0 $}
	
	To compute the entanglement entropy, we first rewrite Eq.~(4) in the main text in the real space representation
	\begin{align}
		|\boldsymbol{k}, \ell, N_b \rangle = \frac{1}{\sqrt{3}} \sum_{m=0,1,2} 
		e^{-i(\frac{2\pi m}{3}\ell - \alpha_m)} 
		\frac{1}{L}\sum_{m_1,m_2=1}^L 
		e^{-(2\pi i/L) (k_1m_1 + k_2m_2)}
		\frac{ (\hat{\psi}_{m_1\boldsymbol{e}_1 + m_2\boldsymbol{e}_2,m}^\dagger )^{N_b}}{\sqrt{N_b!}} |0\rangle
	\end{align}
	The full density matrix for each FBS is defined as $ \rho = |\boldsymbol{k},\ell,N_b\rangle \langle \boldsymbol{k}, \ell, N_b| $, and in the real space representation
	\begin{align}\label{eq:denmat}
		\rho &= \frac{1}{3L^2}\sum_{m,m'=0,1,2} e^{i(\frac{2\pi(m-m')}{3}\ell - (\alpha_m - \alpha_{m'}))} \sum_{m_1,m_2,m_1',m_2'=1}^L
		e^{\frac{2\pi i}{L}(k_1(m_1-m_1') + k_2(m_2-m_2')) }
		\left(
		\frac{ (\hat{\psi}_{m_1\boldsymbol{e}_1 + m_2\boldsymbol{e}_2,m}^\dagger )^{N_b}}{\sqrt{N_b!}} |0\rangle
		\langle 0 | \frac{ (\hat{\psi}_{m_1'\boldsymbol{e}_1 + m_2'\boldsymbol{e}_2,m'} )^{N_b}}{\sqrt{N_b!}}
		\right)
	\end{align}
	Then, we obtain the reduced density matrix by enclosing $ N_s $ subsystem unit cells ($ 3N_s $ sites), dubbed region $ A $, among totally $ 3L^2 $ sites. The remaining part is denoted as $ B $. Due to the form of Eq.~\eqref{eq:denmat}, $ \rho_A $ only involves tracing over configurations of $ 0 $ or $ N_b $ bosons in region $ B $,
	\begin{align}\nonumber
		\rho_A = \text{Tr}_B (\rho) &=
		\langle 0_B| \rho |0_B\rangle + 
		\sum_{m''_1,m''_2, m''\in B} \langle 0_B| \frac{ (\hat{\psi}_{m''_1\boldsymbol{e}_1 + m''_2\boldsymbol{e}_2,m''} )^{N_b}}{\sqrt{N_b!}} \rho \frac{ (\hat{\psi}_{m''_1\boldsymbol{e}_1 + m''_2\boldsymbol{e}_2,m''}^\dagger )^{N_b}}{\sqrt{N_b!}} |0_B \rangle
		\\ \nonumber
		&=
		\frac{1}{3L^2} 
		\sum_{\scriptsize
			\begin{array}{l}
				m_1,m_2,m,\\
				m_1',m_2',m'
			\end{array}\in A} 
		e^{i(\frac{2\pi(m-m')}{3}\ell - (\alpha_m - \alpha_{m'}))}
		e^{\frac{2\pi i}{L}(k_1(m_1-m_1') + k_2(m_2-m_2')) }
		\left(
		\frac{ (\hat{\psi}_{m_1\boldsymbol{e}_1 + m_2\boldsymbol{e}_2,m}^\dagger )^{N_b}}{\sqrt{N_b!}} |0_A \rangle
		\langle 0_A | \frac{ (\hat{\psi}_{m_1'\boldsymbol{e}_1 + m_2'\boldsymbol{e}_2,m'} )^{N_b}}{\sqrt{N_b!}}
		\right) \\
		&\quad + 
		\frac{3L^2-3N_s}{3L^2} |0_A\rangle \langle 0_A|.
	\end{align}
	Here, the first term  would allow for arbitrary indices $ (m_1,m_2,m), (m_1',m_1',m')\in A $ because they all correspond to zero particles in region $ B $. For the second term, there must be $ (m_1,m_2,m) = (m_1',m_2',m') $ equaling to the trace indices $ (m_1'',m_2'',m'')\in B $, so all the phase factors vanish, and there are $ 3L^2 - 3N_s $ sites in region $ B $ giving rise to the prefactor. Denote 
	\begin{align}
		F_{m_1, m_2, m} = e^{-i( \frac{2\pi m }{3}\ell - \alpha_{m})} e^{-\frac{2\pi i}{L} (k_1  m_1 + k_2 m_2)}, 
		\qquad
		|F\rangle = (F_{0,0,0}, F_{0,0,1}, F_{0,0,2}, F_{0,1,0},\dots, F_{L_x^{(A)}, L_y^{(A)}, 2} )^T,
	\end{align}
	the reduced density operator, written in the matrix form, has 
	\begin{align}\label{eq:detail4}
		\rho_A &= 
		\frac{1}{3L^2}
		\begin{pmatrix}
			\begin{pmatrix}
				|F\rangle \langle F|
			\end{pmatrix}_{3N_s\times 3N_s}
			& 0 \\
			0 & 3L^2 - 3N_s
		\end{pmatrix} 
	\end{align}
	Apparently, there are only 2 nonzero eigenvalues. The first one is for the $ (3N_s\times 3N_s) $ matrix whose eigenvector is $ (1/\sqrt{3N_s})|F\rangle $ (note $ \langle F|F\rangle = 3N_s $) corresponding to the eigenvalue $ N_s/L^2 $. The second one is the $ 1\times 1 $ part obviously corresponding to eigenvalue $ (L^2-N_s)/L^2 $. That yields the entanglement entropy
	\begin{align}
		S_{\text{ent}} = -\text{Tr}\left(\rho_A \ln \rho_A \right) 
		= -\gamma\ln\gamma - (1-\gamma)\ln(1-\gamma), \qquad
		\gamma\equiv \frac{N_s}{L^2}.
	\end{align}
	Therefore, for the choices of region $ A $ in the main text Fig.~2 (b), we have $ S_{\text{ent}} = \ln2 \approx 0.6931 $ for $ L=2 (\gamma=1/2) $, and $ S_{\text{ent}} = (4/9)\ln(4/9) + (5/9)\ln(5/9) \approx 0.6870 $ for $ L=3 (\gamma=4/9) $. They lead to slight differences of $ S_{\text{ent}} $ at $ \lambda\rightarrow0 $ due to different subsystem portions $ \gamma=N_s/L^2 $. However, the reference scar vanishing point $ \lambda_0\approx0.135 $ is unlikely to be dominated by such differences, as significant deviations of $ S_{\text{ent}} $ already takes place there compared with $ S_{\text{ent}} $ at $ \lambda\rightarrow0 $. This is also confirmed by the IPR scaling in Fig.~3 (d) of main texts (irrelevant of subsystem size) that also gives $ \lambda_0\approx0.135 $.

	\section{More details for experimental proposals}
	
\begin{figure}
		[h]
		\parbox[b]{2.5cm}{
			\includegraphics[width=2.5cm]{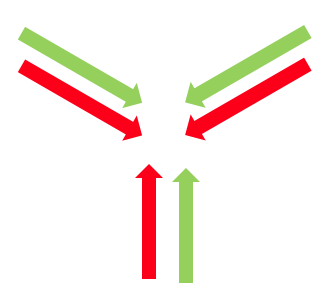}\\
			\quad\\\quad\\
			(a) Lasers
		}
		\parbox[b]{5.9cm}{
			\includegraphics[width=5.5cm]{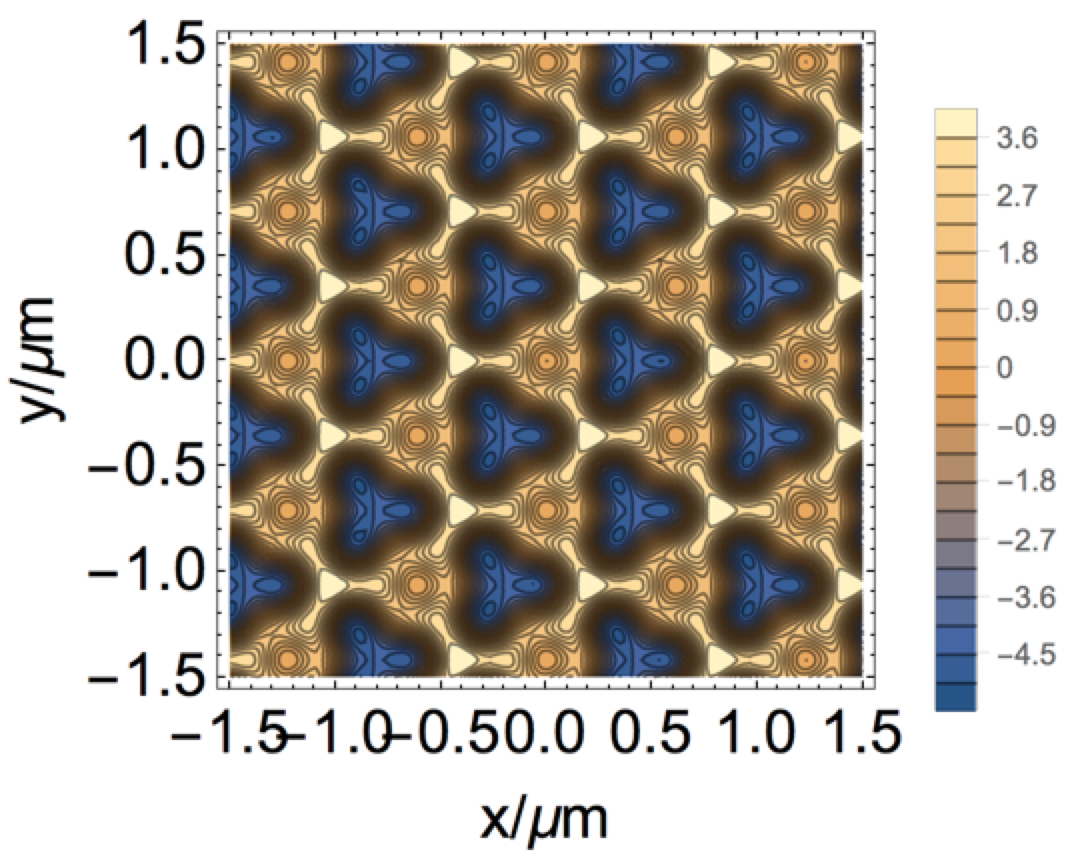}\\
			(b) $ V_0^{(R)} = V_{0}^{(G)} $ }
		\parbox[b]{5.9cm}{
			\includegraphics[width=5.3cm]{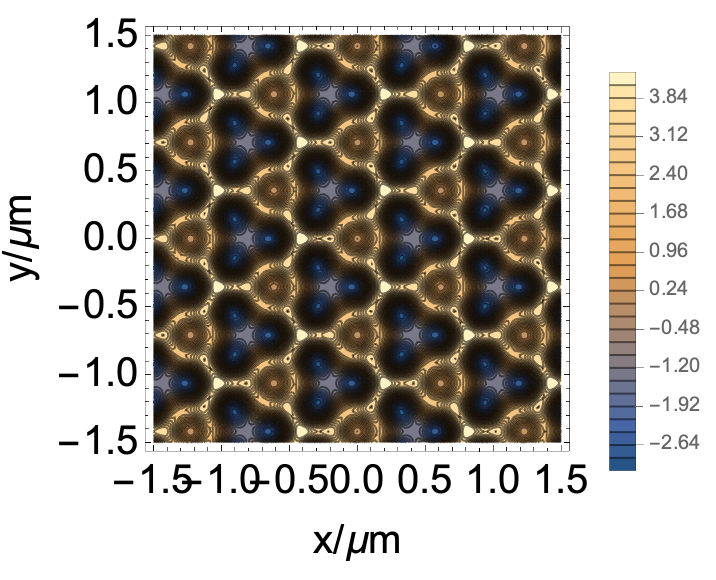}\\
			(b) $ V_0^{(R)} = 0.8 V_{0}^{(G)} $ }
		\caption{\label{fig:tri_kagome}Schemes for the driven trimerized kagome lattice, where $ x_0 \approx 405 \text{nm}, y_0=0 $. The extent of trimerization can be tuned continuously by the relative strength between red and green laser beams.}
	\end{figure}

	This section gives a more detailed account for the experimental proposals. They are most relevant to the kagome lattice platform at Berkeley~\cite{Thomas2017,Barter2020,Leung2020,Brown2021,Jo2012}, while similar schemes can be generalized into other lattices. The laser system we consider consists of 6 beams, with 3 being red ($ \lambda_R=1064 $nm) and 3 green ($ \lambda_G=532 $nm). They are directed along the same plane, where each set of monochromatic laser beams form $ 120^{\circ} $ angles with respect to each other, as in Fig.~\ref{fig:tri_kagome} (a). The laser system exhibits good tunability in forming different lattices in a unified setting, including the (trimerized) kagome, honeycomb, stripe, and Su-Schrieffer-Heeger types of lattices. We would discuss two exemplary choices of experimental setup in the following.

	\subsection{Scheme I: Trimerized kagome lattice (TKL)}
	
	The TKL setting in Ref.~\cite{Barter2020} uses all 6 bichromatic lasers beams, where green beams are polarized in-plane, while red ones are along $ z $. The lattice potential is given by
	\begin{align}\nonumber
		V(x,y) &= V_R(x,y) + V_G(x,y),\\ \nonumber
		V_R(x,y) &=
		- V_0^{(R)} \left|\sum_{j=1}^3  e^{\frac{2\pi i}{\lambda_R} ((x-x_0)\cos\Phi_j + (y-y_0)\sin\Phi_j)} \right|^2 ,
		\\ \nonumber
		V_G(x,y) &= V_0^{(G)} \left(
		+ \left|  \sum_{j=1}^3 \cos(\Phi_j+\pi/2)e^{\frac{2\pi i }{\lambda_G}  (x\cos\Phi_j + y\sin\Phi_j)} \right|^2 
		+ \left|  \sum_{j=1}^3 \sin(\Phi_j+\pi/2)e^{\frac{2\pi i }{\lambda_G}  (x\cos\Phi_j + y\sin\Phi_j)} \right|^2 \right) , \\
		&
		(\Phi_1,\Phi_2,\Phi_3) = \left(
		-\frac{\pi}{2}, \frac{\pi}{6}, \frac{5\pi}{6}
		\right).
	\end{align}
	
	TKLs correspond to our main text model, and the associated driving protocols are
	\begin{itemize}
		\item 
		Shake the lattice circularly in order to endow a $ \pi/2 $ flux per triangle. This shaking is kept on throughout the whole period.
		\item
		Drive the lattice potential strength periodically, such that the Hamiltonian switches between hopping-dominant terms for $ \hat{H}_1 $ and Hubbard interaction (plus possible sublattice energy offset) dominant terms for $ \hat{H}_2 $. Then, the Hamiltonians (in ideal situations) take the form as in Eq.~(1) of the main text.
		\item 
		For specific parameter control, one can fix the duration $ t_1 $ for the first half of a period according to $ \phi_1 = Jt_1/\hbar \approx 2\pi/3\sqrt3 $, where $ J $ is the hopping strength of strong bonds. Then, $ \lambda = J'/J $, with $ J' $ the hopping strength for weak bonds. Similarly, the ``interaction strength" for the Floquet parameter $ \phi_2 = Ut_2/\hbar $ can be controlled by the duration $ t_2 $ in the second half of a period, where $ U $ is the Hubbard interaction strength. One driving period $ T = t_1 + t_2 $ here.
	\end{itemize}

	\subsection{Scheme II: Shaken honeycomb lattice}
\begin{figure}
		[h]
		\parbox[b]{3cm}{\includegraphics[width=3cm]{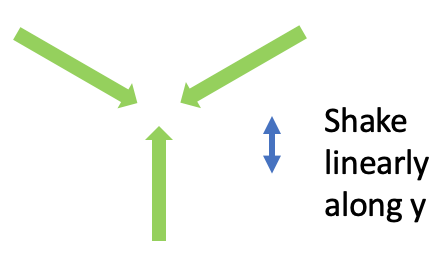}\\
			\qquad\\ \quad \\
			(a) Lasers and shaking scheme}
		\parbox[b]{5.5cm}{\includegraphics[width=5.5cm]{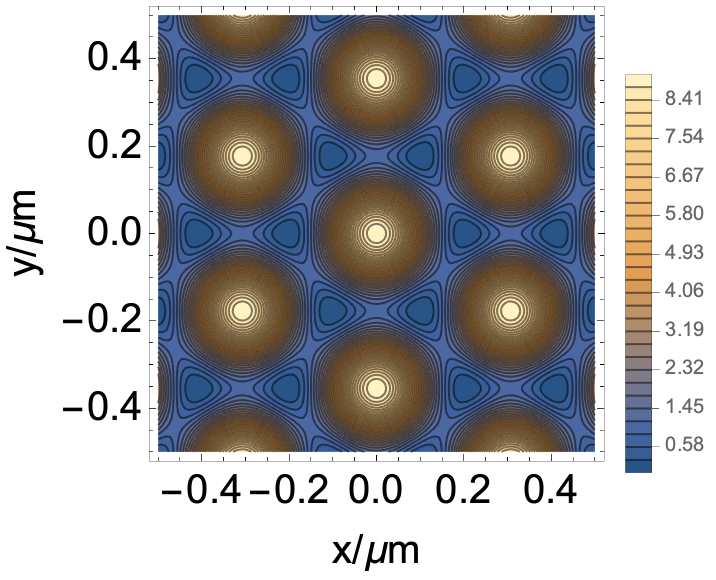}\\
			(b) Potential contours}
		\parbox[b]{2.5cm}{\includegraphics[width=2.5cm]{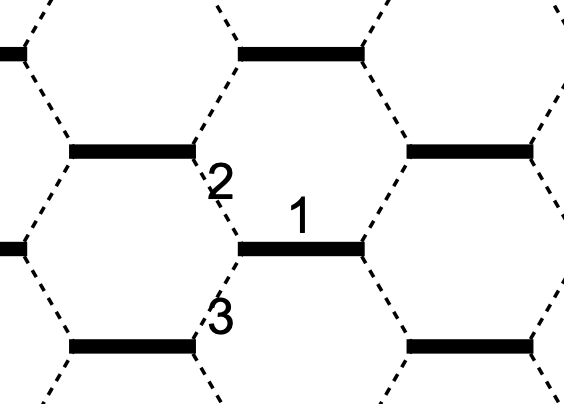}\\ 
			(c) Honeycomb lattice with dimerized hopping strength}
		\parbox[b]{6cm}{
			Three nearest neighbor bonds
			$ \boldsymbol{d}_1 = \boldsymbol{e}_x, 
			\boldsymbol{d}_2 = -\frac{1}{2}\boldsymbol{e}_x + \frac{\sqrt3}{2}\boldsymbol{e}_y,
			\boldsymbol{d}_3 = -\frac{1}{2}\boldsymbol{e}_x - \frac{\sqrt3}{2}\boldsymbol{e}_y $
			\\
			$ \gamma_m \rightarrow \gamma_m J_0 \left(
			\frac{m\omega \boldsymbol{d}_m\cdot \boldsymbol{e}_y}{\hbar}
			\right)  $\\ 
			Strong bond $ 1 $ not affected, weak bonds $ \rightarrow0 $ when shaking frequency $ \frac{m\omega\sqrt3}{2\hbar} \rightarrow 2.405  $
			\\
			\includegraphics[width=4cm]{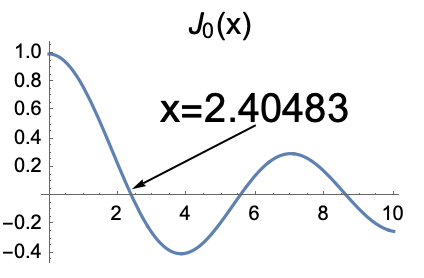}
		}
		\caption{\label{fig:dim_honeycomb} The dimerized honeycomb lattice. The extent of dimerization is achieved by tuning the shaking frequencies. $ J_0(x) $ is the Bessel function of the first kind.}
	\end{figure}

	The honeycomb lattice~\cite{Brown2021} only requires monochromatic lasers, while for our purposes the lattice should be dimerized as described later. Here, we propose to use the green laser beams to generate a honeycomb lattice, while the red beams would be used later to engineer the initial state. So the green beams here should be polarized along $ z $, giving
	\begin{align}
		V(x,y) = V_G'(x,y) = +V_0 \left|\sum_{j=1}^3  e^{\frac{2\pi i}{\lambda_G} (x\cos\Phi_j + y\sin\Phi_j)} \right|^2 ,
		\qquad		
		(\Phi_1,\Phi_2,\Phi_3) = \left(
		-\frac{\pi}{2}, \frac{\pi}{6}, \frac{5\pi}{6}
		\right).
	\end{align} 
	The lattice potentials are illustrated in Fig.~\ref{fig:dim_honeycomb}. 
	The corresponding driving protocols here are
	\begin{enumerate}
		\item 
		One can smoothly control the extent of dimerization, shown in Fig.~\ref{fig:dim_honeycomb} (c) (d), by shaking linearly the whole lattice at all time~\cite{Quelle2017}.
		
		\item 
		Similar to the trimerized kagome lattice, we can add a driving in terms of laser intensity to produce the relatively slow Floquet driving, where the first and second step of a Floquet driving produce the Hamiltonians
		\begin{align}\label{eq:hn1}
			\text{Hopping to three neighbors with unequal strength: \qquad}  &\frac{\hat{H}_1T}{2\hbar} = \phi_1\sum_{\boldsymbol{r}} \hat{\psi}^\dagger_{\boldsymbol{r}} 
			(
			\hat{\psi}_{\boldsymbol{r}+\boldsymbol{d}_1} 
			+ \lambda
			(\hat{\psi}_{\boldsymbol{r}+\boldsymbol{d}_2} + \hat{\psi}_{\boldsymbol{r}+\boldsymbol{d}_3}  )
			)
			\\\label{eq:hn2}
			\text{Onsite interactions and possible sublattice energy offsets:\qquad}
			& 
			\frac{\hat{H}_2T}{2\hbar} = \phi_2 \sum_{\boldsymbol{r}} n_{\boldsymbol{r}}(n_{\boldsymbol{r}}-1) + \theta_{\boldsymbol{r}} n_{\boldsymbol{r}}
		\end{align}
	\end{enumerate}
	
	\subsection{Initial state preparation}
\begin{figure}
		[h]
		\parbox{6cm}{
			\includegraphics[width=6cm]{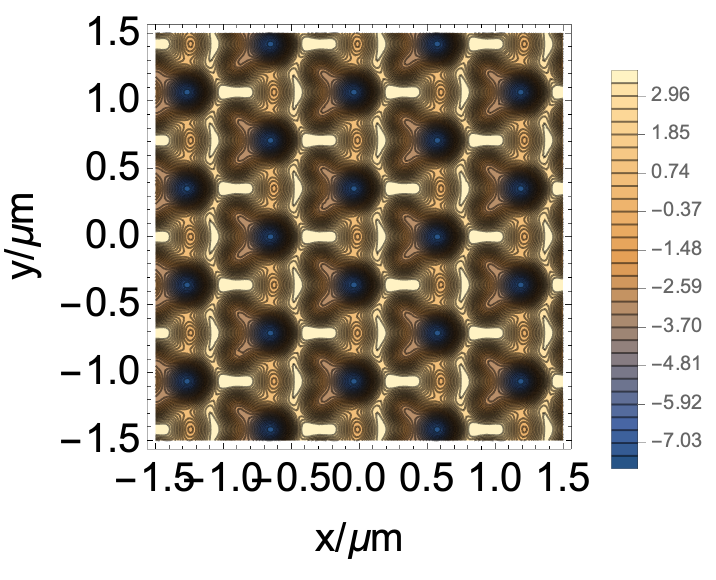}\\
			(a) Trimerized kagome case}
		\parbox{6cm}{
			\includegraphics[width=6cm]{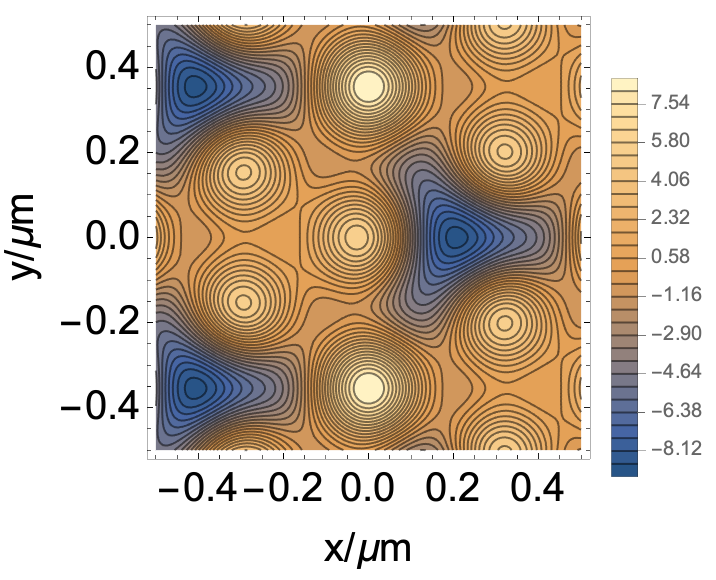} \\
			(b) Dimerized honeycomb case}
		\caption{\label{fig:ini}Laser schemes to prepare for initial states. (a) For the trimerized kagome lattice case, the scheme is realized by slightly modifying the setting in Fig.~\ref{fig:tri_kagome} by tuning the relative phase of the red beam, such that $ x_0 \approx 0.5nm $ and the minima of the red laser potential matches one sublattice site. After loading the atoms to one sublattice as in (a), one could quench $ x_0\rightarrow 0.405nm $ and recover the scheme in Fig.~\ref{fig:tri_kagome}. (b) For the dimerized honeycomb case, in addition to the scheme in Fig.~\ref{fig:dim_honeycomb} using purely green beams, one could further apply the red beams polarized along $ z $ forming triangular lattices. Matching the potential minima to one of the honeycomb site would produce the plotted potentials for preparing initial states. }
	\end{figure}

	Now, we show the scheme to realize initial states of depositing particles in one sublattice. That can be achieved by taking advantage of the highly tunable kagome optical lattice platform, for both the trimerized kagome and dimerized honeycomb settings, as illustrated in Fig.~\ref{fig:ini}.
	
	\subsection{Detection of sublattice particle number by band mapping}
	
	The projection measurement of particle number in individual sublattice sites of a non-primitive unit cell of an optical lattice can be performed in three steps. First, the lattice depth is suddenly increased to quench tunneling between lattice sites and project the subsystem in each site to approximately a number Fock state. Then, the superlattice potential is adiabatically deformed, say by changing the relative position of the two underlying sublattices that add up to create the trimerized kagome lattice or the dimerized honeycomb lattice, to energetically detune all subllatice sites. If the detuning is sufficiently large and all the sublattice sites are decoupled, then each energy band of the system is predominantly associated with one sublattice site only. Finally, band mapping, a standard technique where the lattice potential is addiabatically turned off to map quasimomentum to free-particle momentum, is performed. Band population can thus be measured in time-of-flight imaging.
	
	Sublattice-site particle number measurement performed with a similar but slightly different technique can be found in \cite{Taie2015}.

	\subsection{Simulations of results}
	To simulate concrete experimental situation with large lattices and filling fractions, we resort to a semiclassical numerical method, the truncated Wigner approximation (TWA)~\cite{Polkovnikov2010}. Roughly speaking, this method goes beyond a mean field analysis by sampling over different initial states
	\begin{align}
		W[\Phi_{\boldsymbol{r}\mu}^{(0)}] = \frac{1}{2\pi\sigma^2} e^{-|\Phi_{\boldsymbol{r}\mu}^{(0)} - \sqrt{n_{\boldsymbol{r}\mu}^{(0)}}|^2/2\sigma^2}
	\end{align}
	where quantum operators $ \hat{\psi}_{\boldsymbol{r}\mu} $ are replaced by their mean field values $ \psi_{\boldsymbol{r}\mu} $, and their could deviation from the initial state values $ \{n_{\boldsymbol{r}\mu}^{(0)} \} $ in different samplings. The fluctuation $ \sigma=1/2 $ means there is on average one half excessive particles per site $ \langle \delta n_{\boldsymbol{r}\mu} \rangle = \int_{-\infty}^\infty |\Phi_{\boldsymbol{r}\mu}^{(0)} - \sqrt{n^{(0)}_{\boldsymbol{r}\mu}} |^2 W[\Phi_{\boldsymbol{r}\mu}^{(0)}] d(Re\Phi_{\boldsymbol{r}\mu}) d(Im\Phi_{\boldsymbol{r}\mu}) = 2\sigma^2 = 1/2 $, which will be canceled by the symmetrization process for transforming operators into Weyl symbols, i.e. $ \hat{n}_{\boldsymbol{r}\mu} = \frac{1}{2}(\hat{\psi}_{\boldsymbol{r}\mu}^\dagger \hat{\psi}_{\boldsymbol{r}\mu} + \hat{\psi}_{\boldsymbol{r}\mu} \hat{\psi}_{\boldsymbol{r}\mu}^\dagger ) - 1/2 \rightarrow |\Phi_{\boldsymbol{r}\mu}|^2 - 1/2 $~\cite{Polkovnikov2010}. 
	Meanwhile, the evolutions are still prescribed by classical differential equations, which can be obtained by first using the Heisenberg's equation of motion $ i\partial_t \hat{\psi}_{\boldsymbol{r}\mu} = [\hat{\psi}_{\boldsymbol{r}\mu},H] $ and then replacing $ \hat{\psi}_{\boldsymbol{r}\mu} $ with Weyl symbols (complex numbers) $ \Phi_{\boldsymbol{r},\mu} $. For instance, the main text model of trimerized kagome lattice in Eq.~(1) prescribes the equations of motion in each period $ T $ as 
	\begin{align}
		t\in[0,T/2): \qquad & \qquad
		\partial_t\Phi_{\boldsymbol{r}\mu} = J \sum_{\nu\ne\mu} f_{\mu\nu} ( \Phi_{\boldsymbol{r}\nu} + \lambda \Phi_{\boldsymbol{r}-\boldsymbol{e}_\nu + \boldsymbol{e}_\mu, \nu} ) ,\\
		t\in[T/2,T): \qquad & \qquad
		i\partial_t \Phi_{\boldsymbol{r}\mu} = 2U |\Phi_{\boldsymbol{r}\mu}|^2 \Phi_{\boldsymbol{r}\mu}
	\end{align}
	Here the parameter $ J, U $ are related to those in Eq.~(1) by $ JT/2\hbar = \phi_1, UT/2\hbar = \phi_2 $. 
	This way, the semiclassical results are expected to capture the quantum fluctuations during early time of evolution, which is most relevant to experimental observations. As a side remark, we notice that if only a pure mean field simulation is adopted (no initial state sampling), one would observe a deceptive infinite time DTC oscillation without decay at all for a rather wide range of initial states. That contradicts exact diagonalization and analytical results, and the generic thermalizing nature confirmed by level spacing statistics. Therefore, it is of vital importance to incorporate fluctuations at least for the initial states so as to simulate a realistic situation at early time.

	\begin{figure}
		[h]
		\parbox[b]{5.5cm}{\includegraphics[width=5.5cm]{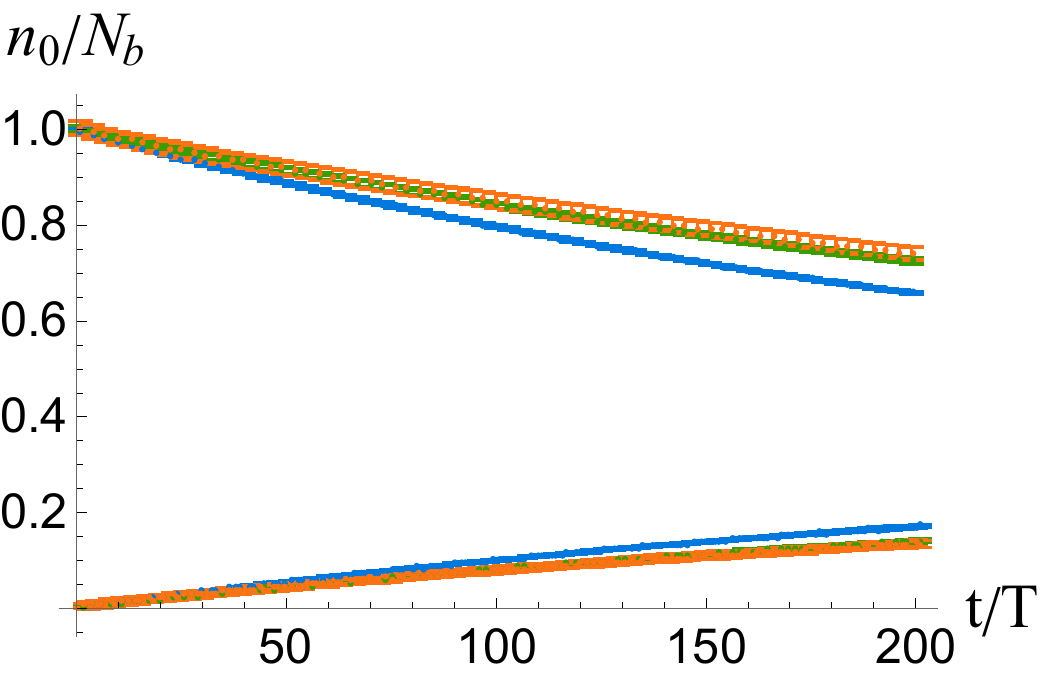} (a) Numerical results}
		\parbox[b]{3.5cm}{\includegraphics[width=3.cm]{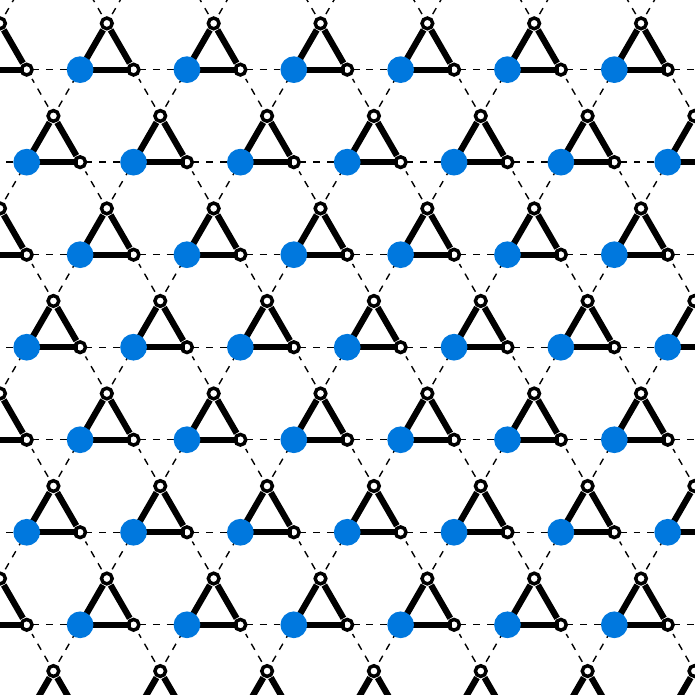} \\ 
			(experimentally accessible)
			(b) No separation}
		\parbox[b]{3.5cm}{\includegraphics[width=3.cm]{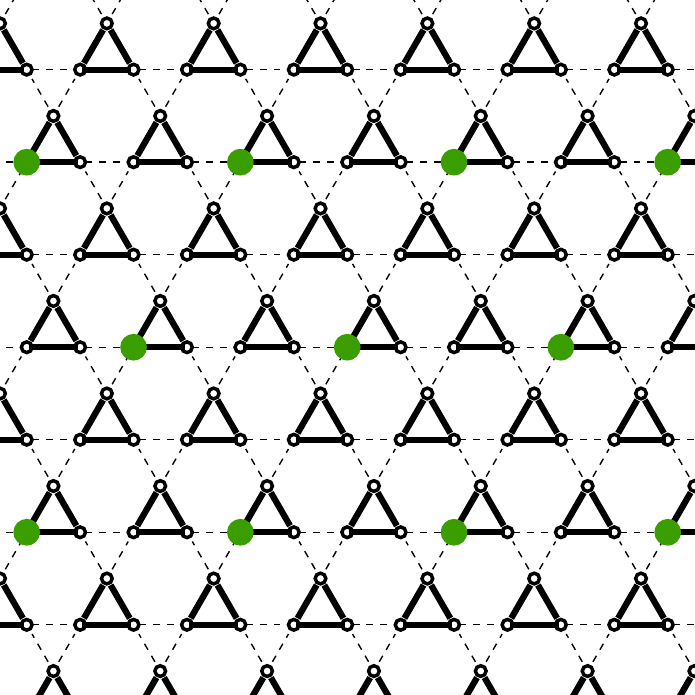} \\\qquad \\
			(c) Separation by 1 cell }
		\parbox[b]{3.5cm}{\includegraphics[width=3.cm]{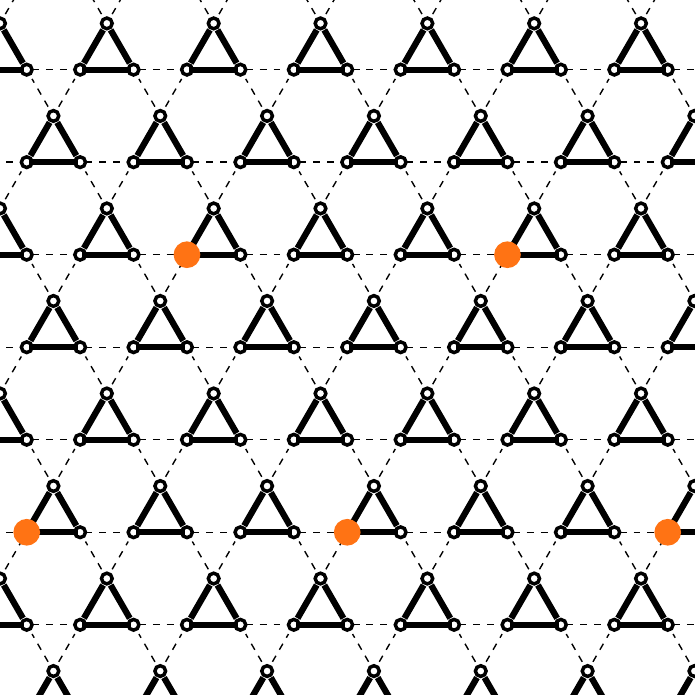} \\ \qquad \\
			(d) Separation by 2 cells}
		\caption{\label{fig:twa_tkl}TWA simulation of scar enforced DTC dynamics in trimerized kagome lattice. We simulate a lattice containing $ 12\times 12 $ unit cells under periodic boundary conditions. Each populated site denoted by colored dots in (b) -- (d) contains 5 bosons as the initial state. Evolutions from these states are simulated in (a) for the corresponding colors. Two sublattices are denoted as $ \mu=0 $ and $ \mu=1 $ following the convention in main text. $ n_0 $ is the total particle number in sublattice $ 0 $, and $ N_b $ the total number of bosons in all sites. Parameters for the model written in Eq.~\eqref{eq:hn1} and \eqref{eq:hn2} are $ \phi_1 = \pi/2, \phi_2 = 1.1, \theta_{\boldsymbol{r}}=0, \lambda=0.05 $. Simulations contain 3000 initial state Monte-Carlo samples for each site and error bars denote the standard deviation when all data are group into 10 bins.}
	\end{figure}
	
	\begin{figure}
		[h]
		\parbox[b]{5.5cm}{\includegraphics[width=5.5cm]{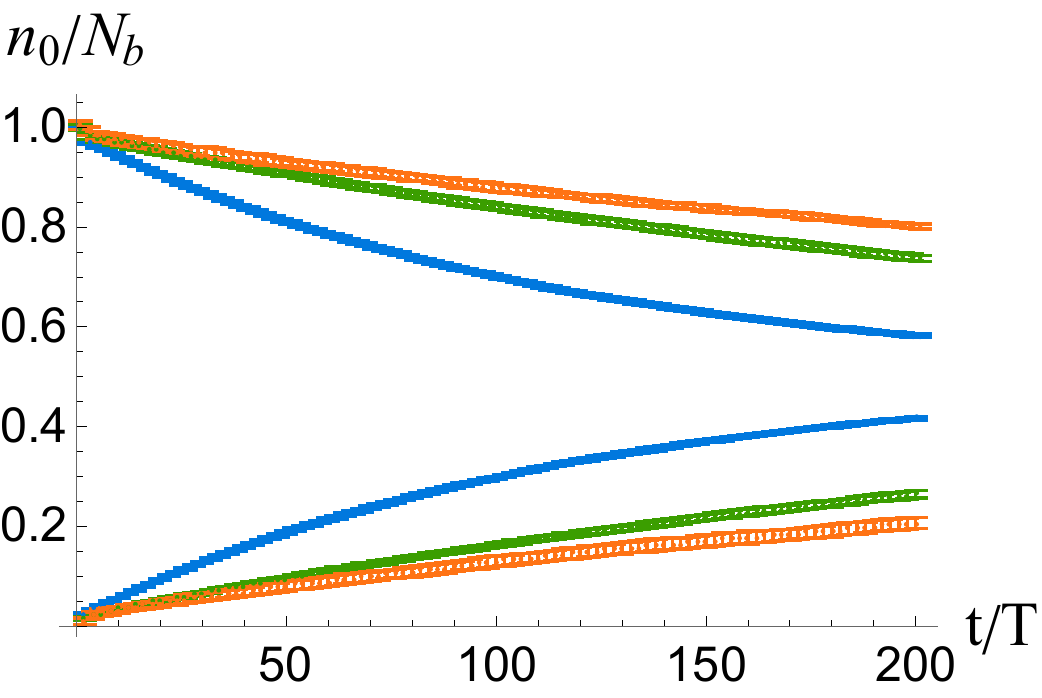} (a) Numerical results}
		\parbox[b]{3.5cm}{\includegraphics[width=3.cm]{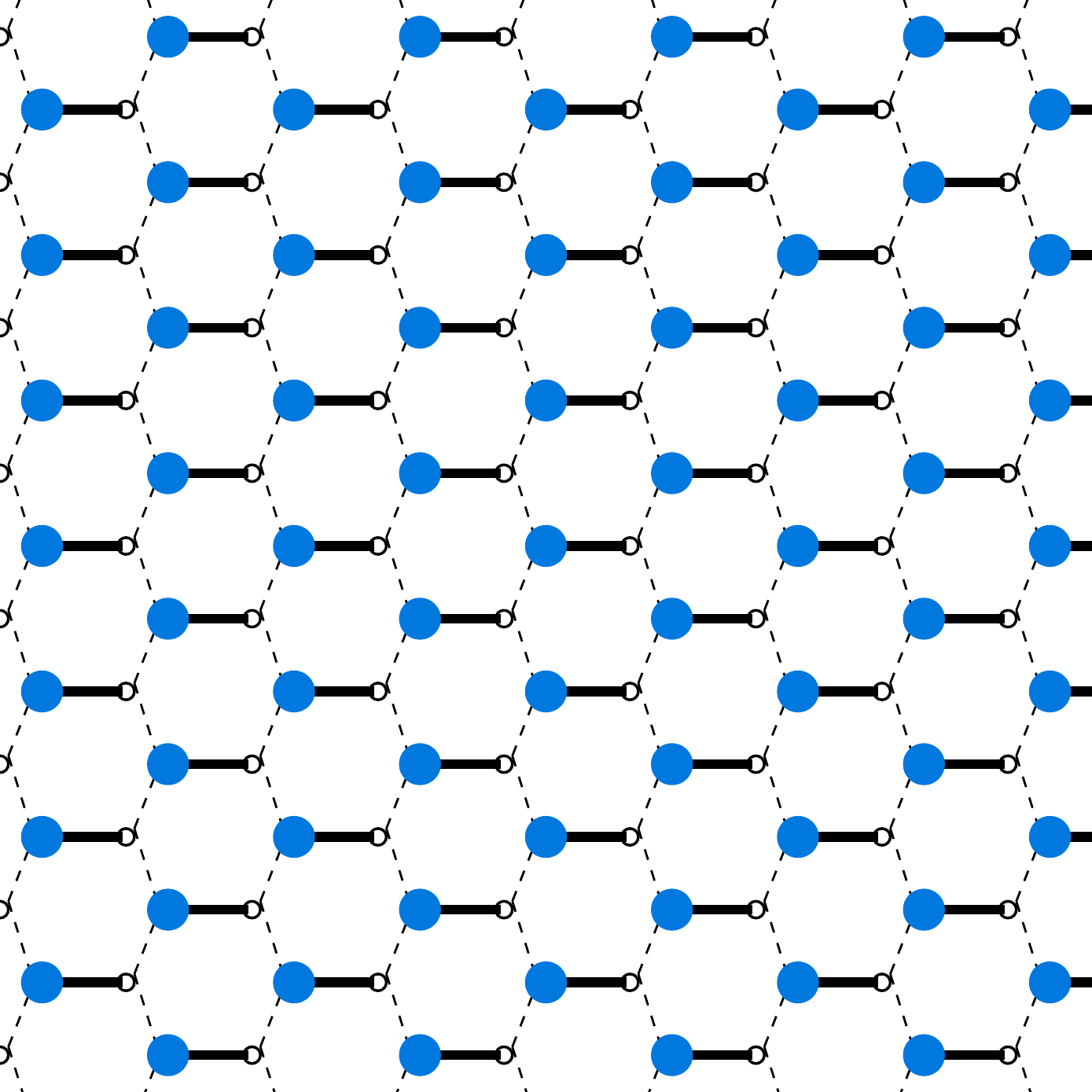} \\ \qquad \\
			(b) No separation}
		\parbox[b]{3.5cm}{\includegraphics[width=3.cm]{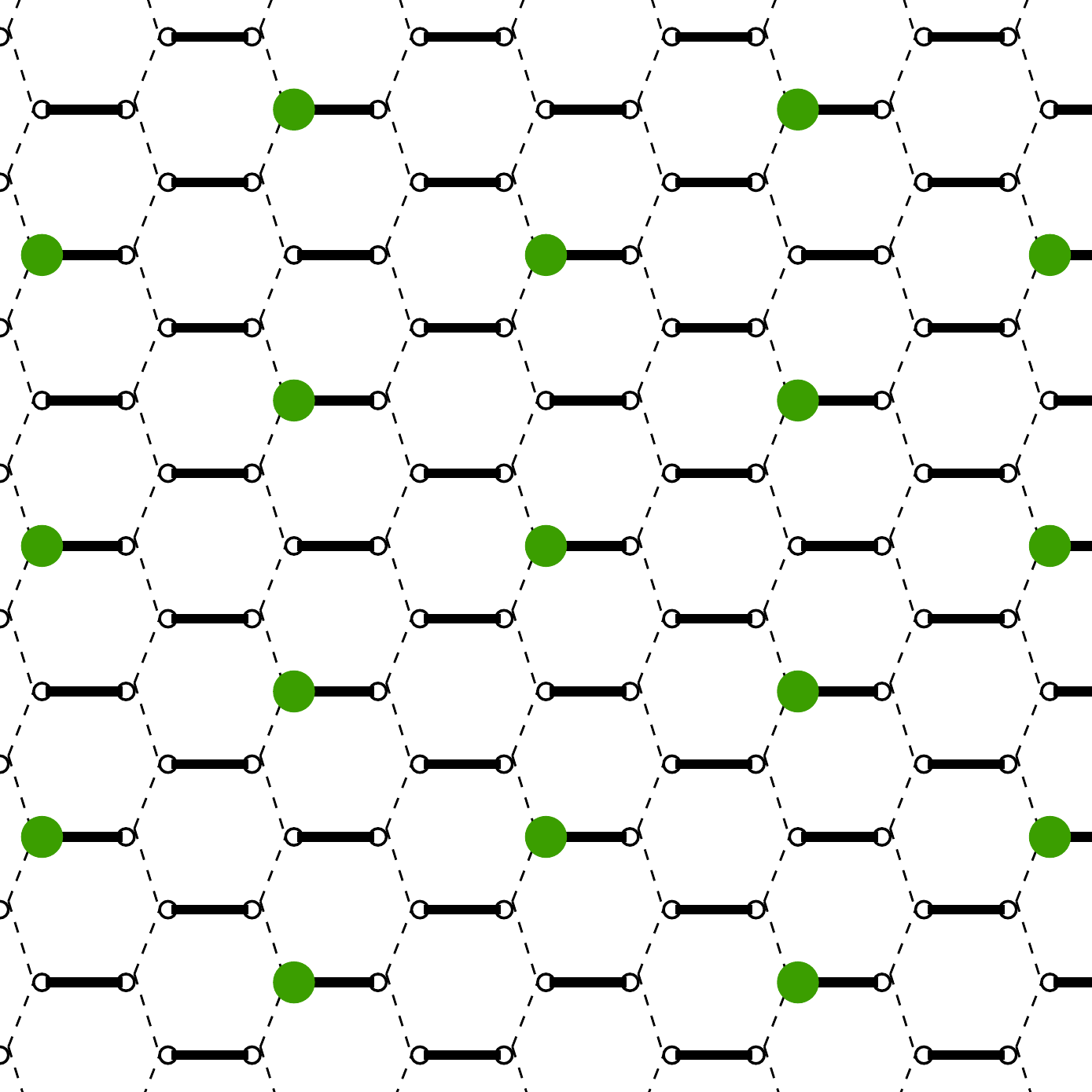} \\
			(experimentally accessible)
			(c) Separation by 1 cell }
		\parbox[b]{3.5cm}{\includegraphics[width=3.cm]{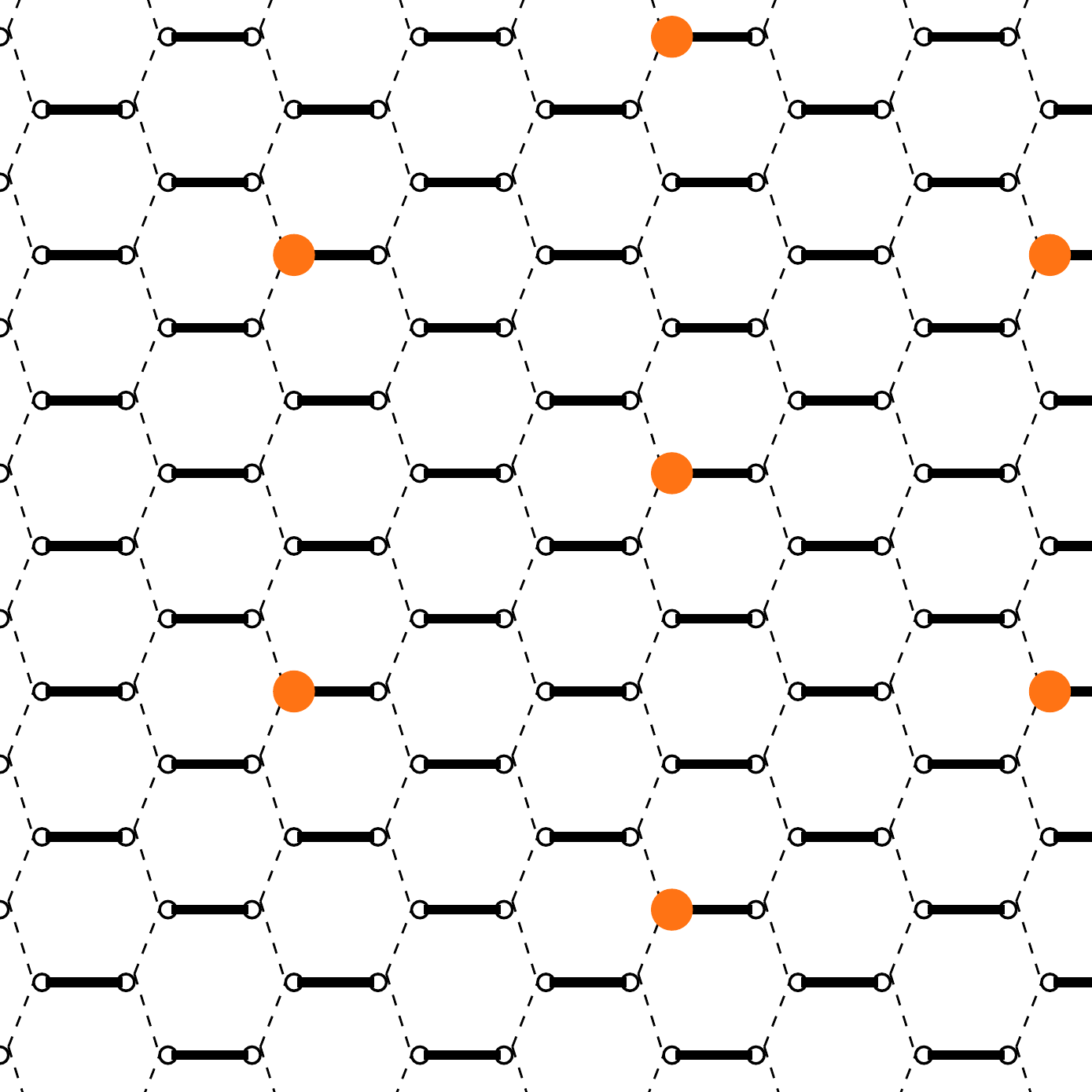} \\ \qquad \\
			(d) Separation by 2 cells}
		\caption{\label{fig:twa_dhl} Similar simulations as in Fig.~\ref{fig:twa_tkl} for the dimerized honeycomb lattice. System size is again $ 12\times 12 $ unit cells, and each initially populated site in (b) -- (d) contains 5 bosons. $ n_0 $ means the total particle number in sublattice $ \mu=0 $ and $ N_b $ is the total boson number. Parameters for the model written in Eq.~\eqref{eq:hn1} and \eqref{eq:hn2} are $ \phi_1 = \pi/2, \phi_2 = 1.1, \theta_{\boldsymbol{r}}=0, \lambda=0.05 $.  Simulations contain 3000 initial state Monte-Carlo samples for each site and error bars denote the standard deviation when all data are group into 10 bins.}
	\end{figure}
	
	The results for sublattice density dynamics is shown in Fig.~\ref{fig:twa_tkl} and \ref{fig:twa_dhl}, where we compare the DTC decay rates starting from different initial states. It is clear that with larger spatial separations for the initially populated sites lead to prolonged oscillations due to longer time needed for FBS localized at different unit cells to interact with each other.

\end{document}